\documentclass[twocolumn,pre,superscriptaddress,10pt,aps,floatfix]{revtex4-2}
\usepackage[english]{babel}
\usepackage{graphicx}
\usepackage{amsmath}
\usepackage[colorlinks,linkcolor=blue,citecolor=blue,urlcolor=blue]{hyperref}
\usepackage{color}
\usepackage{soul}
\definecolor{dblue} {RGB}{28,130,185}
\let\oldaddcontentsline\addcontentsline
\newcommand{\stoptocentries}{\renewcommand{\addcontentsline}[3]{}}
\newcommand{\starttocentries}{\let\addcontentsline\oldaddcontentsline}
\definecolor{nred}{RGB}{224,0,0}
\definecolor{nblue}{RGB}{28,130,185}

\begin{document}
\title{Restoring ergodicity in a strongly disordered interacting chain}
\author{B. Krajewski}
\affiliation{Department of Theoretical Physics, Faculty of Fundamental Problems of Technology, \\ Wroc\l aw University of Science and Technology, 50-370 Wroc\l aw, Poland}
\author{L. Vidmar}
\affiliation{Department of Theoretical Physics, J. Stefan Institute, SI-1000 Ljubljana, Slovenia}
\affiliation{Department of Physics, Faculty of Mathematics and Physics, University of Ljubljana, SI-1000 Ljubljana, Slovenia\looseness=-1}
\author{J. Bon\v ca}
\affiliation{Department of Physics, Faculty of Mathematics and Physics, University of Ljubljana, SI-1000 Ljubljana, Slovenia\looseness=-1}
\affiliation{Department of Theoretical Physics, J. Stefan Institute, SI-1000 Ljubljana, Slovenia}
\author{M. Mierzejewski}
\affiliation{Department of Theoretical Physics, Faculty of Fundamental Problems of Technology, \\ Wroc\l aw University of Science and Technology, 50-370 Wroc\l aw, Poland}
\date{\today}
\begin{abstract}
We  consider a chain of interacting fermions with random disorder that was intensively  studied in the context of many-body localization. 
We show that only a small fraction of the two-body interaction represents a true local perturbation to the Anderson insulator.
While this true perturbation is nonzero at any finite disorder strength $W$, it  decreases with increasing $W$.
This establishes a view that the strongly disordered system should be viewed as a weakly perturbed integrable model, i.e., a weakly perturbed Anderson insulator. As a consequence, the latter can hardly be distinguished from a strictly integrable system in finite-size calculations at large $W$.
We then introduce a rescaled model in which the true perturbation is of the same order of magnitude as the other terms of the Hamiltonian, and show that the system remains ergodic at arbitrary large disorder.
\end{abstract}
\maketitle
\stoptocentries
%================================================================================

{\it Introduction.} The interplay between disorder and interactions in quantum systems has recently attracted significant interest.
Some of the most exciting ideas were formulated within the framework of many-body localization (MBL), which is conjectured to be as a phase of matter that violates ergodicity in spite of presence of interactions~\cite{basko06, gornyi05, oganesyan07, Rahul15, altman15, alet_laflorencie_18, abanin2019}. 

The disordered  systems exhibit several unusual properties, in particular extremely slow dynamics 
\cite{znidaric08,serbyn13_1,kjall14,bera15,lev15,serbyn2017,sels2020,vidmar2021}
that was frequently interpreted as a precursor to MBL 
\cite{luitz2016prl,luitz116,znidaric16,gopal17,kozarzewski18,prelovsek217,new_karrasch,prelovsek2018a}. However, one of the the most important questions about MBL is related to its stability in the thermodynamic limit.
Until recently, the results of essentially all studies in one-dimensional (1D) spin-1/2 systems with disorder were interpreted in terms of a stable MBL phase~\cite{znidaric08, pal10, monthus10, khatami_rigol_12, bardarson12, serbyn13_1, deluca13, kjall14, bera15, luitz15, torres15, serbyn15, lev15, devakul_singh_15, serbyn_moore_16, bertrand_garcia_16, mierzejewski2016, steinigeweg16, filippone_brouwer_16, khemani_lim_17, khemani_sheng_17, detomasi_bera_17, pietracaprina_parisi_17, gray_bose_18, sierant_zakrzewski_19, schiulaz_torresherrera_19, mace_alet_19, roy_chalker_19, roy_logan_20, tarzia_20}.
Recent work has, however, highlighted robustness of ergodicity at moderate disorder~\cite{suntajs_2020}, which may eventually suggest that stability of MBL may not be taken for granted.
Signatures of robustness of ergodicity were also reported in several subsequent works~\cite{suntajs_bonca_20a, kieferemmanouilidis_unanyan_20, sels2020, kieferemmanouilidis_unanyan_21, leblond_sels_21, vidmar2021, Sels_dilute_2021} and they triggered, among others, activities to gain a better insight into the avalanche theory of ergodicity breaking transitions~\cite{deroeck_huveneers_17, luitz_huveneers_17, thiery_huveneers_18, crowley_chandran_20, Sels_2022, Morningstar2022, suntajs_vidmar_22, crowley_chandran_22b}.
However, many recent numerical studies  are interpreted in terms of existence of a stable MBL phase ~\cite{Panda2020, Sierant2020, sierant_lewenstein_20, abanin_bardarson_21, corps_molina_21, prakash_pixley_21, schliemann_costa_21, hopjan_orso_21, solorzano_santos_21, detomasi_khaymovich_21, crowley_chandran_22, ghosh_znidaric_22, bolther_kehrein_22, yintai_yufeng_22, sierant2022, Morningstar2022, sutradhar_ghosh_22, trigueros_cheng_22, shi_khemani_22}. Then, 
the MBL-to-thermal phase transition may occur at much stronger disorders than suggested by earlier numerical calculations \cite{Morningstar2022}.

Motivated by these open questions, it is an outstanding problem to understand why exact numerical studies can give rise to formulation of contradictory expectations for the same models in the thermodynamic limit.
More generally, what are the crucial ingredients of interacting systems with disorder that make identification of their key physical properties so challenging?

This Letter provides new perspective into studies of robustness of ergodicity and its detection in finite systems. 
For the model of interacting spinless fermions with disorder, which is mappable onto the paradigmatic random-field Heisenberg chain, we show that only a small fraction of the two-body interaction represents a true local perturbation to the Anderson insulator. 
The true perturbation decreases with increasing the disorder, and it eventually becomes too weak to be captured by finite-size numerical calculations.
Consequently, the strongly disordered system should be viewed as a weakly perturbed Anderson insulator.
As an application of this insight, we introduce a rescaled model in which the strength of the true perturbation matches the energy density of the Anderson insulator.
We argue that the latter model remains ergodic at essentially any finite disorder, and show that the matrix elements of observables are consistent with the eigenstate thermalization hypothesis (ETH)~\cite{deutsch_91, srednicki_94, rigol_dunjko_08, dalessio_kafri_16}.
	
{\it Set-up.}
We study interacting fermions in a 1D disordered lattice with $L$ sites and periodic boundary conditions. The system is described by the Hamiltonian $H=H_0+H_{\Delta}$, referred to as the {\it standard model} further on. 
The first term describes the Anderson insulator,
\begin{eqnarray}
 H_0&=&\sum_{i=1}^L h_{i},  \label{h0}  \\
h_i&=&\frac{1}{2} (a^{\dagger}_{i+1} a_i + {\rm H.c.})+ \frac{\epsilon_i}{2} (n_i-\frac{1}{2}) +  \frac{ \epsilon_{i+1}}{2} (n_{i+1}-\frac{1}{2}),  \nonumber
\end{eqnarray}
where $\epsilon_i$ is a random potential with box distribution, $-W \le \epsilon_i \le W$, $a^{\dagger}_i$ creates a spinless fermion at site $i$ and $n_i=a^{\dagger}_{i} a_i $. The second  term  is the two-body interaction,
\begin{equation}
H_{\Delta}=\Delta \sum_{i=1}^{L}  N_i, \quad \quad N_i= \left(n_i-\frac{1}{2}\right) \left(n_{i+1}-\frac{1}{2}\right),  \label{hd} 
\end{equation}
where we take $\Delta=1$ so that $H$ can be mapped onto to the widely studied random-field Heisenberg model. 
The noninteracting part (i.e., the integrable part) of the Hamiltonian is diagonal in the Anderson basis  
\begin{equation}
H_0= \sum_{\alpha} \varepsilon_{\alpha} Q_{\alpha} +{\rm const}, \quad \quad   Q_{\alpha}=2 a^{\dagger}_{\alpha}a_{\alpha} -1, \label{h0diag}
\end{equation} 
where  $a_{\alpha}=\sum_{i} u^{*}_{i \alpha} a_i$ and  $u_{i \alpha}= \langle i | \alpha \rangle$ are components of the single-particle wavefunction of the Anderson state $\alpha$. 

\begin{figure}[!tb]
\includegraphics[width=1.0\columnwidth]{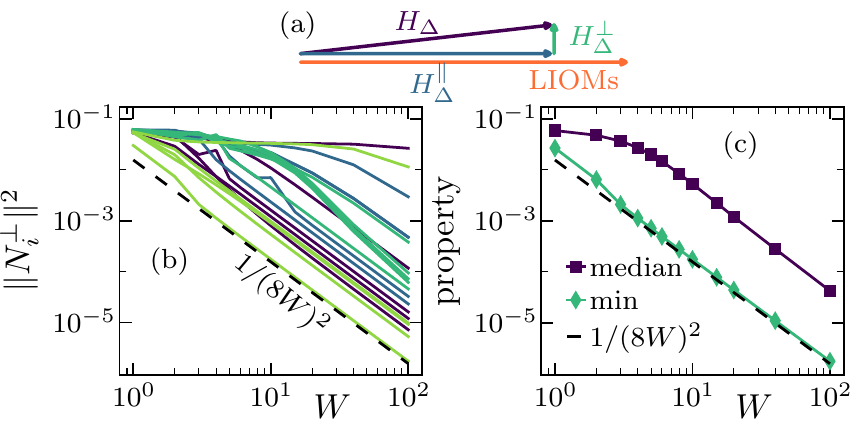}
\caption{(a) Sketch of the construction in Eq.~(\ref{hsplit}).
(b) Dependence of $||N^{\perp}_i ||^2 $ on $W$, where various curves of the same color correspond to different $i$ but the same disorder realization
 [we keep $\epsilon_i/W$=const when increasing $W$]. Various colors correspond to different realizations of disorder.
(c) Two statistical properties of $||N^{\perp}_i ||^2$ from  $10^4$ curves as those in (b): median and minimum.
Dashed line is the lower bound $1/(8W)^2$, see~\cite{suppmat}.
Results in (b,c) are obtained at $L=14$ and $L/2$ fermions. 
}
\label{fig1}
\end{figure}

As a central step of our approach, we split  the interaction term in Eq.~(\ref{hd})  into two orthogonal parts,
\begin{equation} \label{hsplit}
H_{\Delta}=H_{\Delta}^{\parallel}+H_{\Delta}^{\perp}\;, 
\;\;\; {\rm with}\;\;\; \langle H_{\Delta}^{\parallel} H_{\Delta}^{\perp} \rangle=0 \;,
\end{equation}
where orthogonality is defined via the Hilbert-Schmidt inner product as $\langle A  B \rangle=\frac{1}{Z} {\rm Tr}(A^{\dagger}  B)$, the trace is carried out over  many-body states and $Z$ is the dimension of the Fock space.  
In Eq.~(\ref{hsplit}), $H_{\Delta}^{\parallel}$ represents a projection of $H_{\Delta}$ onto local integrals of motion of the Anderson insulator, thus $[H_{\Delta}^{\parallel},H_0]=0$.
As a consequence, we identify the interaction in $H^{\perp}_{\Delta}$ as a {\it true perturbation} to the Anderson insulator, and we argue that it represents a local Hamiltonian.
The idea of our approach is sketched in Fig.~\ref{fig1}(a).
Here, locality of operators (e.g., $h_i$ or $N_i$) refers to the size of their support in real space which is fixed and does not grow with $L$. Linear combinations of the latter operators (e.g., $H_0$ or $H_{\Delta}$) are also considered as local.

Below we show that the squared norm of the true perturbation, $||H^{\perp}_{\Delta}||^2$, decays asymptotically for large $W$ as $1/W^2$, whereas the squared norm of the Anderson model,  $||H_0||^2$, grows as $W^2$. 
Then, for sufficiently large $W$, the perturbation appears to be too weak to break integrability of a finite system.
Here, the squared norms of observables are defined  as $ ||A ||^2= \langle A  A \rangle$.

{\it Local integrals of motion.}
The traceless operators $Q_\alpha$ from  Eq.~(\ref{h0diag}) represent the one-body local integrals of motion of the Anderson insulator. We briefly refer to them as LIOMs.
We sort them according to the maxima of the single-particle wave-functions, $u_{i \alpha}$, i.e., we find $i_\alpha= {\rm max}_i |u_{i \alpha}|$ and sort them such that $i_{\alpha} \le i_{\alpha'} $ for $\alpha \le \alpha'$. Roughly speaking, for open boundary conditions the Anderson states with  $\alpha \ll L$ are localized at the left edge of the system whereas the states with $\alpha \sim L$ are localized at the right edge. 
Importantly, a remarkable property of the Anderson insulator is that not only the LIOMs $Q_{\alpha}$ are local, but so are also their products,
$Q^{(2)}_{\alpha,d} \equiv Q_{\alpha} Q_{\alpha+d}$, provided that the distance $d=1,...,d_{max}$ is small  compared to $L$ and $d_{max}$ does not grow with the system size~\footnote{In contrast, in translationally invariant integrable models the products of local charges are typically nonlocal thus, in the thermodynamic limit, they have no projection on local observables.}.
We briefly refer to these $Q^{(2)}_{\alpha,d}$ as two-body LIOMs.
 
It is straightforward to show that $H_{\Delta}$ from Eq.~(\ref{hd}) has no projection on traceless LIOMs $Q_\alpha$, see~\cite{suppmat} for details.
Therefore, we introduce an operator $N_i^{\parallel}$ that is a linear combination of two-body LIOMs, such that
\begin{equation}
N_i^{\parallel}=\sum_{d=1}^{d_{max}} \sum_{\alpha=1}^{L} \langle Q^{(2)}_{\alpha,d} N_i \rangle   Q^{(2)}_{\alpha,d},\quad \quad N_i^{\perp}=N_i-N_i^{\parallel}\;. \label{proj}
\end{equation}
The operator $N_i^{\parallel}$ can be interpreted as a projection  of a local interaction onto two-body LIOMs, and hence it corresponds to an interaction that does not break integrability of the Anderson insulator.
In contrast, $N_i^{\perp}$ can be viewed as the true perturbation.

We stress two important technical details.
First, we only consider results for  $d_{max} =2$ in this Letter, whereas in~\cite{suppmat} we show that additional contributions coming from  
$d_{max} > 2$ are negligible at strong disorder.
Second, in the Fock space that consists of $2^L$ many-body configurations, the occupations of LIOMs $Q_\alpha$ are independent and their products $Q^{(2)}_{\alpha,d}$ are mutually orthogonal and normalized, i.e., $\langle Q^{(2)}_{\alpha,d} Q^{(2)}_{\alpha',d'} \rangle=\delta_{\alpha,\alpha'} \delta_{d,d'} $. As a consequence, Eq.~(\ref{proj}) represents an orthogonal projection for which $\langle N_i^{\parallel}  N_i^{\perp} \rangle=0$. 
However, the actual calculations are carried out in a subspace with $L/2$ fermions, in which the LIOMs are not independent since $\sum_{\alpha}Q_{\alpha}=0$, and their products are not traceless since $\langle Q^{(2)}_{\alpha,d} \rangle = O(1/L)$. Then,  
one needs to reorthogonalize the set of $Q^{(2)}_{\alpha,d}$, as explained in~\cite{suppmat}. 

{\it Norm of the true perturbation.}
We can now express $H^{\parallel}_{\Delta}$ and $H^{\perp}_{\Delta}$ from Eq.~(\ref{hsplit}) using Eq.~(\ref{proj}) as 
\begin{equation} \label{hpp}
H^{\parallel}_{\Delta}=\Delta \sum_{i=1}^{L}  N_i^{\parallel} \;\;\; {\rm and} \;\;\; H^{\perp}_{\Delta}=\Delta \sum_{i=1}^{L}  N_i^{\perp} \;.
\end{equation}
Since Eq.~(\ref{proj}) assures locality of $N_i^{\parallel}$ and  $N_i^{\perp}$, then $H^{\parallel}_{\Delta}$ and $H^{\perp}_{\Delta}$ are also local as they are defined as linear combinations of local operators $N_i$ and $Q^{(2)}_{\alpha,d}$. 
The physical meaning of $H^{\perp}_{\Delta}$ can be understood by inspecting the identity (see~\cite{suppmat} for a derivation)
\begin{equation} \label{def_Hperpnorm}
||H^{\perp}_{\Delta}||^2=||H_{\Delta}||^2 -\sum_{\alpha,d} \langle H_{\Delta} Q^{(2)}_{\alpha,d}\rangle^2,
\end{equation}
which shows that the more two-body LIOMs $Q^{(2)}_{\alpha,d}$ one takes, the smaller is the norm of $H^{\perp}_{\Delta}$. 
Indeed, the essence of our approach is a systematic elimination of local contributions to $H_{\Delta}$ which commute with the integrable Hamiltonian $H_0$.

%--------------------------------------------------------------------------------
%----------------------------------------------------------------------------

Figures~\ref{fig1}(b) and~\ref{fig1}(c) study the dependence on $W$ of the squared norms $|| N^{\perp}_i ||^2$ that contribute to the norm of $H^{\perp}_{\Delta}$ in Eq.~(\ref{hpp}).
Each curve in Fig.~\ref{fig1}(b) is obtained for a single site $i$ and a single realization of disorder, while Fig.~\ref{fig1}(c) shows the median
and the minimum of $10^4$ curves as those in Fig.~\ref{fig1}(b).
One observes huge fluctuations between various sites and disorder realizations.
Nevertheless, at sufficiently large $W$ all curves eventually decay as  
$||N^{\perp}_i ||^2 \propto 1/W^2$, see Fig.~\ref{fig1}(b). 
For strong disorder we establish an $L$-independent bound $|| N^{\perp}_i ||^2 \ge 1/(8W)^2$, which accurately reproduces the numerical results in Fig.~\ref{fig1}(c) already at $W >3$. The derivation of the bound and the $L$-dependence of $|| N^{\perp}_i ||^2$ are discussed in~\cite{suppmat}.

Summarizing this part, we stress that the perturbation to the Anderson insulator is not determined by the entire interaction term but rather by the projected operators, $N_i^{\perp}$.  
This perturbation becomes very weak at strong disorder, $|| N^{\perp}_i ||  \sim 1/W$, but remains nonzero for arbitrary finite $W$. Obviously, such a small but non-vanishing perturbation poses a challenge for finite-size numerical calculations.

{\it Ergodicity in the rescaled model.} 
We complement the above analysis by introducing a model in which the norm of the true perturbation does not vanish with increasing $W$.
To this end we study the {\it rescaled model} Hamiltonian
\begin{equation}
\tilde{H} =\sum_i h_i + \sum_i  \frac{|| h_i || }{|| N^{\perp}_i ||} N^{\perp}_i\;, \label{h2}
\end{equation}
where $h_i$ denotes the local term (the energy density operator) of the Anderson model from Eq.~(\ref{h0}) and $ N^{\perp}_i$ represents the density of the true perturbation from Eq.~(\ref{hpp}).
Both energy density operators $h_i$ and $N_i$ are defined on the link between sites $i$ and $i+1$.

The rescaled model~(\ref{h2}) associates the strength of the perturbation with the strength of the disorder. 
In particular, the energy density of the true perturbation, cf.~the second term on the r.h.s.~of Eq.~(\ref{h2}), equals to the energy density of the Anderson insulator, for which the squared norm is $|| h_i ||^2=(2+ \epsilon^2_i+ \epsilon^2_{i+1})/16$.
In the standard model, this roughly corresponds to the regime
$\Delta \propto W$,
for which one may expect an ergodic-to-nonergodic transition.
(The nonergodic phase is conjectured to be reentrant as a function of the interaction strength, see, e.g., Fig.~1 in~\cite{lev15}.)
Below we explore robustness of ergodicity in the rescaled model~(\ref{h2}).

\begin{figure}[!t]
\includegraphics[width=1.0\columnwidth]{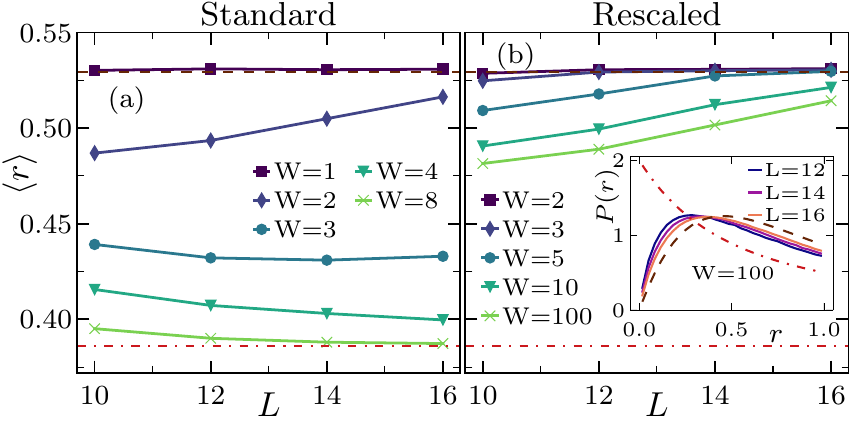}
\caption{Average gap ratio, $\langle r \rangle$, at various  $L$ and $W$ calculated in  subspaces with $L/2$ fermions for (a) the standard model $H$ from Eqs.~(\ref{h0})-(\ref{hd}) and (b) the rescaled model $\tilde{H}$ from Eq.~(\ref{h2}). The averaging is carried out over $Z/3$  levels from the middle of the spectrum and over 4000 realizations of disorder.
Inset in (b): probability density function $P(r)$ in the rescaled model at $W=100$ and various $L$. Dash-dotted and dashed lines show the analytical predictions for the Poisson distribution ~\cite{oganesyan07} and the GOE~\cite{atas2013,fremling2022}, respectively (see also~\cite{suppmat}). }
\label{fig2}
\end{figure}
 
As a simple test of ergodicity we  study the average ratio of nearest level spacings $\langle r \rangle$ (i.e., the  gap ratio), see~\cite{suppmat} for a definition.
The results are shown in Figs.~\ref{fig2}(a) for the standard model $H$ from Eqs.~(\ref{h0})-(\ref{hd}) and in Fig.~\ref{fig2}(b)  for the rescaled model $\tilde{H}$ from Eq.~(\ref{h2}).
In the standard model the results clearly deviate from the value $r\simeq 0.53$ in the Gaussian orthogonal ensemble (GOE) already at $W \gtrsim 3$, which was observed in many previous studies, see, e.g., Refs.~\cite{luitz15}.
However,  the rescaled model remains ergodic at essentially all disorders, provided that the system is sufficiently large.  As an additional test,  we determine a distribution of $r$ without any averaging, i.e., via collecting results from different disorder realizations as well as different eigenstates (from the middle third of spectra). 
The inset of Fig.~\ref{fig2}(b) shows the resulting probability density function $P(r)$ at various $L$. 
A comparison with  analytical results~\cite{oganesyan07, atas2013, alet2022, fremling2022, suppmat} confirms that at large $L$ the results approach the GOE prediction even at $W=100$.

\begin{figure}[!htb]
\includegraphics[width=1.0\columnwidth]{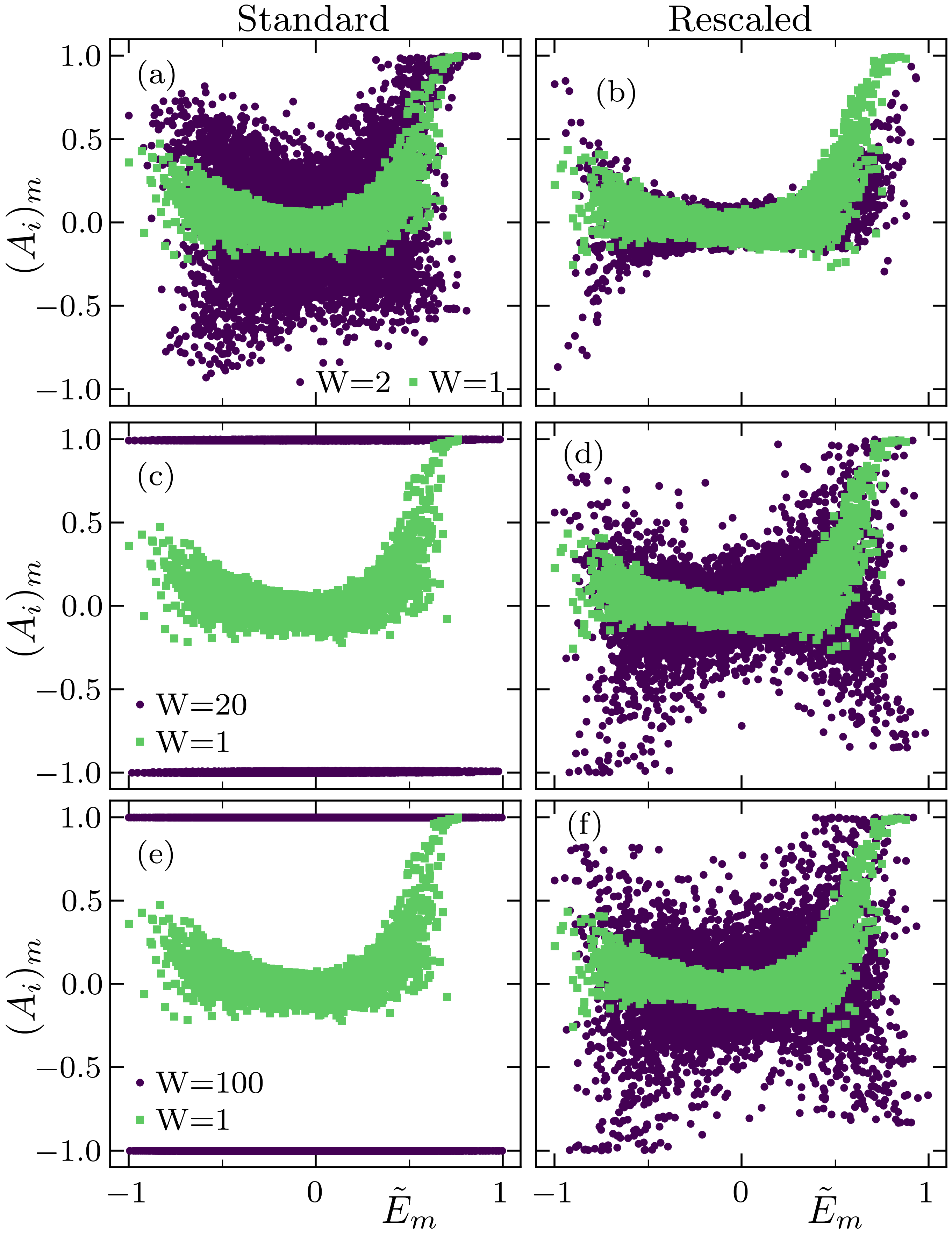}
\caption{Diagonal matrix elements $(A_i)_m= \langle E_m | A_i |E_m \rangle$, where $A_i=2n_i-1$, at $L=16$ and different $W$. 
Results are shown for a single site $i$ and a single realization of disorder.
(a,c,e) The standard model $H$ from Eqs.~(\ref{h0})-(\ref{hd}) and (b,d,f) the rescaled model $\tilde{H}$ from Eq.~(\ref{h2}).
We rescale the energies as $\tilde{E}_m=E_m/|E_0|$, where $E_0$ is the ground state energy.
}
\label{fig3}
\end{figure}
 
\begin{figure}[!htb]
\includegraphics[width=1.0\columnwidth]{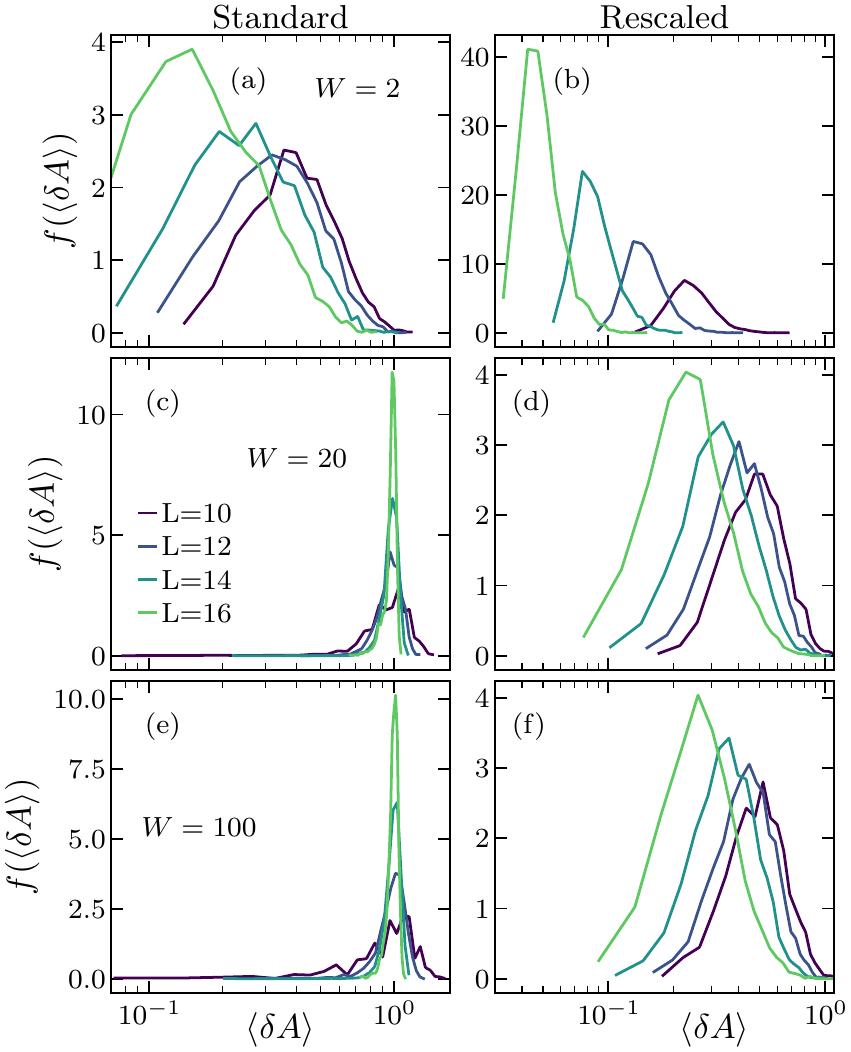}
\caption{
Probability density function $f$ of the eigenstate-to-eigenstate fluctuations $\langle \delta A \rangle$ from Eq.~(\ref{eefluct}) for various $L$. The distributions are calculated at single lattice site and different realizations of disorder for 
(a,c,e) the standard model $H$ and (b,d,f) the rescaled model $\tilde{H}$. 
}
\label{fig4}
\end{figure}
  
{\it ETH analysis.} Finally, we test ergodicity of the rescaled Hamiltonian by studying  the ETH. As observables we consider site occupations $ A_i=2n_i-1$.
Note that a linear combination of $A_i$, the imbalance  $I=\sum_i (-1)^i A_i$,  has been commonly studied in the context of ergodic-nonergodic transition and is accessible in cold-atom experiments~\cite{schreiber15}. Following a standard procedure~\cite{rigol_dunjko_08}, we calculate the diagonal matrix elements  $(A_i)_m=\langle E_m| A_i |E_m \rangle $ where $|E_m \rangle$ are the many-body eigenstates of either the standard Hamiltonian $H$, or the rescaled Hamiltonian  $\tilde{H}$, see Fig.~\ref{fig3}.
In a finite system described by the standard model, one observes  $(A_i)_m=\pm 1$ at all energies at strong disorder, see Figs.~\ref{fig3}(c) and~\ref{fig3}(e), and hence the ETH appears to be violated, suggesting nonergodic behavior. However, in the rescaled model the fluctuations of matrix elements are rather modest even at extremely strong disorder $W = 100$, see Fig.~\ref{fig3}(f).

To study fluctuations of the diagonal matrix elements we calculate the average eigenstate-to-eigenstate fluctuations~\cite{Kim_strong2014, Mondaini2016, jansen_stolpp_19},
\begin{equation} \label{eefluct}
\langle \delta A \rangle= 1/{\cal{Z}} \sum_m |(A_i)_{m+1}-(A_i)_{m}| \;,
\end{equation}
where the averaging is carried out over ${\cal{Z}}=Z/5$ states from the middle of the many-body spectrum. 
Figure~\ref{fig4}  shows the probability density functions, $f(\langle \delta A \rangle)$, calculated at a single lattice site and different disorder realizations, for both the standard and the rescaled model.
%(see also~\cite{suppmat} for the distributions of $\langle \delta A \rangle$).
In the standard model one obtains $\langle \delta A \rangle \simeq 1$ at large disorder [cf.~Figs.~\ref{fig4}(c) and~\ref{fig4}(e)], and the absence of any visible  $L$-dependence of the distributions may be interpreted as a violation of the ETH. In the rescaled model the distribution of $\langle \delta A \rangle$ is rather broad for the accessible system sizes.
Nevertheless $\langle \delta A \rangle$ appears to decay with $L$ suggesting $\langle \delta A \rangle \to 0$ in the thermodynamic limit. Due to the width of the distributions, one cannot unambiguously confirm exponential decay of the latter quantity.
However, such a decay is strongly suggested by the decay of the median, see also~\cite{suppmat}.
 
{\it Conclusions.}
The main goal of this Letter was to identify the origin of complexity that emerges in the numerical studies of ergodicity in interacting fermions subject to random disorder.
We showed that the two-body interaction term  $H_{\Delta}$~(\ref{hd})  cannot be considered as a perturbation to the  Anderson insulator $H_0$~(\ref{h0}) since only a small fraction of the two-body interaction, denoted as $H_{\Delta}^{\perp}$, does not commute with $H_0$.
We referred to the latter as the true local perturbation and we showed that its relative norm decays with disorder as $||H_{\Delta}^{\perp} ||/|| H_0|| \sim W^{-2}$. 
On the other hand, the norm is also bounded from below so it remains nonzero at large but finite $W$. 
It is then clear that the interpretation of finite-size numerical calculations at large $W$ is challenging  since finite integrable systems with small perturbations are hardly distinguishable from strictly integrable systems.

It appears that the two regimes in which interpretation of numerical results has rather low ambiguity are: the regime of small and moderate $W$, for which robustness of ergodicity was already established, and the regime where the strength of the true local perturbation $H_{\Delta}^{\perp}$ is rescaled.
Here we considered the latter scenario and introduced a rescaled model in which the energy density of the perturbation equals that of the Anderson insulator.
Studying the short-range level statistics and the ETH indicators in the rescaled model we showed that ergodicity persists up to extremely strong disorders, such as $W = 100$. 

While focusing on 1D interacting fermions with random disorder, the main idea of our approach can be applied to an arbitrary model in any dimension.
In particular, the method of identifying the true local perturbation allows for an unambiguous classification of the perturbation strength, and hence provides a new perspective into distinction between weakly and strongly perturbed integrable systems. 
Systems of broad interest to which the method can straightforwardly be applied in the near future are interacting fermions subject to quasiperiodic~\cite{iyer_oganesyan_13} or linear~\cite{schulz_hooley_19, nieuwenburg_baum_19} potentials.
 
\acknowledgements 
We acknowledge discussions with  M. Rigol. We acknowledge the support by the National Science Centre, Poland via project 2020/37/B/ST3/00020 (B.K and M.M.), the support by the Slovenian Research Agency (ARRS), Research Core Fundings Grants P1-0044 (L.V. and J.B.) and J1-1696 (L.V.).
 
\bibliographystyle{biblev1}
\bibliography{ref_mbl}

\begin{thebibliography}{10}
\expandafter\ifx\csname url\endcsname\relax
  \def\url#1{{\tt #1}}\fi
\expandafter\ifx\csname urlprefix\endcsname\relax\def\urlprefix{URL }\fi
\expandafter\ifx\csname bibinfo\endcsname\relax\def\bibinfo#1#2{#2}\fi
\expandafter\ifx\csname eprint\endcsname\relax\def\eprint#1{\url{#1}}\fi

\bibitem{basko06}
\bibinfo{author}{D.~Basko}, \bibinfo{author}{I.~Aleiner}, and
  \bibinfo{author}{B.~Altshuler}, \bibinfo{title}{Metal--insulator transition
  in a weakly interacting many-electron system with localized single-particle
  states},
  \bibinfo{journal}{\href{http://dx.doi.org/10.1016/j.aop.2005.11.014}{Ann.
  Phys.}} \href{http://dx.doi.org/10.1016/j.aop.2005.11.014}{{\bf
  \bibinfo{volume}{321}}, \bibinfo{pages}{1126}}
  (\href{http://dx.doi.org/10.1016/j.aop.2005.11.014}{\bibinfo{year}{2006}}).

\bibitem{gornyi05}
\bibinfo{author}{I.~V. Gornyi}, \bibinfo{author}{A.~D. Mirlin}, and
  \bibinfo{author}{D.~G. Polyakov}, \bibinfo{title}{Interacting electrons in
  disordered wires: Anderson localization and low-$t$ transport},
  \bibinfo{journal}{\href{http://dx.doi.org/10.1103/PhysRevLett.95.206603}{Phys.
  Rev. Lett.}} \href{http://dx.doi.org/10.1103/PhysRevLett.95.206603}{{\bf
  \bibinfo{volume}{95}}, \bibinfo{pages}{206603}}
  (\href{http://dx.doi.org/10.1103/PhysRevLett.95.206603}{\bibinfo{year}{2005}}).

\bibitem{oganesyan07}
\bibinfo{author}{V.~Oganesyan} and \bibinfo{author}{D.~A. Huse},
  \bibinfo{title}{Localization of interacting fermions at high temperature},
  \bibinfo{journal}{\href{http://dx.doi.org/10.1103/PhysRevB.75.155111}{Phys.
  Rev. B}} \href{http://dx.doi.org/10.1103/PhysRevB.75.155111}{{\bf
  \bibinfo{volume}{75}}, \bibinfo{pages}{155111}}
  (\href{http://dx.doi.org/10.1103/PhysRevB.75.155111}{\bibinfo{year}{2007}}).

\bibitem{Rahul15}
\bibinfo{author}{R.~Nandkishore} and \bibinfo{author}{D.~A. Huse},
  \bibinfo{title}{Many-body-localization and thermalization in quantum
  statistical mechanics},
  \bibinfo{journal}{\href{http://dx.doi.org/10.1146/annurev-conmatphys-031214-014726}{Ann.
  Rev. Cond. Mat. Phys.}}
  \href{http://dx.doi.org/10.1146/annurev-conmatphys-031214-014726}{{\bf
  \bibinfo{volume}{6}}, \bibinfo{pages}{15}}
  (\href{http://dx.doi.org/10.1146/annurev-conmatphys-031214-014726}{\bibinfo{year}{2015}}).

\bibitem{altman15}
\bibinfo{author}{E.~Altman} and \bibinfo{author}{R.~Vosk},
  \bibinfo{title}{Universal dynamics and renormalization in many-body-localized
  systems},
  \bibinfo{journal}{\href{http://dx.doi.org/10.1146/annurev-conmatphys-031214-014701}{Annu.
  Rev. Condens. Matter Phys.}}
  \href{http://dx.doi.org/10.1146/annurev-conmatphys-031214-014701}{{\bf
  \bibinfo{volume}{6}}, \bibinfo{pages}{383}}
  (\href{http://dx.doi.org/10.1146/annurev-conmatphys-031214-014701}{\bibinfo{year}{2015}}).

\bibitem{alet_laflorencie_18}
\bibinfo{author}{F.~Alet} and \bibinfo{author}{N.~Laflorencie},
  \bibinfo{title}{{Many-body localization: An introduction and selected
  topics}},
  \bibinfo{journal}{\href{http://dx.doi.org/10.1016/j.crhy.2018.03.003}{C. R.
  Physique}} \href{http://dx.doi.org/10.1016/j.crhy.2018.03.003}{{\bf
  \bibinfo{volume}{19}}, \bibinfo{pages}{498}}
  (\href{http://dx.doi.org/10.1016/j.crhy.2018.03.003}{\bibinfo{year}{2018}}).

\bibitem{abanin2019}
\bibinfo{author}{D.~A. Abanin}, \bibinfo{author}{E.~Altman},
  \bibinfo{author}{I.~Bloch}, and \bibinfo{author}{M.~Serbyn},
  \bibinfo{title}{Colloquium: Many-body localization, thermalization, and
  entanglement},
  \bibinfo{journal}{\href{http://dx.doi.org/10.1103/RevModPhys.91.021001}{Rev.
  Mod. Phys.}} \href{http://dx.doi.org/10.1103/RevModPhys.91.021001}{{\bf
  \bibinfo{volume}{91}}, \bibinfo{pages}{021001}}
  (\href{http://dx.doi.org/10.1103/RevModPhys.91.021001}{\bibinfo{year}{2019}}).

\bibitem{znidaric08}
\bibinfo{author}{M.~\v{Z}nidari\v{c}}, \bibinfo{author}{T.~Prosen}, and
  \bibinfo{author}{P.~Prelov\ifmmode~\check{s}\else \v{s}\fi{}ek},
  \bibinfo{title}{Many-body localization in the {Heisenberg XXZ} magnet in a
  random field},
  \bibinfo{journal}{\href{http://dx.doi.org/10.1103/PhysRevB.77.064426}{Phys.
  Rev. B}} \href{http://dx.doi.org/10.1103/PhysRevB.77.064426}{{\bf
  \bibinfo{volume}{77}}, \bibinfo{pages}{064426}}
  (\href{http://dx.doi.org/10.1103/PhysRevB.77.064426}{\bibinfo{year}{2008}}).

\bibitem{serbyn13_1}
\bibinfo{author}{M.~Serbyn}, \bibinfo{author}{Z.~Papi\'{c}}, and
  \bibinfo{author}{D.~A. Abanin}, \bibinfo{title}{Universal slow growth of
  entanglement in interacting strongly disordered systems},
  \bibinfo{journal}{\href{http://dx.doi.org/10.1103/PhysRevLett.110.260601}{Phys.
  Rev. Lett.}} \href{http://dx.doi.org/10.1103/PhysRevLett.110.260601}{{\bf
  \bibinfo{volume}{110}}, \bibinfo{pages}{260601}}
  (\href{http://dx.doi.org/10.1103/PhysRevLett.110.260601}{\bibinfo{year}{2013}}).

\bibitem{kjall14}
\bibinfo{author}{J.~A. Kj{\"{a}}ll}, \bibinfo{author}{J.~H. Bardarson}, and
  \bibinfo{author}{F.~Pollmann}, \bibinfo{title}{Many-body localization in a
  disordered quantum {Ising} chain},
  \bibinfo{journal}{\href{http://dx.doi.org/10.1103/PhysRevLett.113.107204}{Phys.
  Rev. Lett.}} \href{http://dx.doi.org/10.1103/PhysRevLett.113.107204}{{\bf
  \bibinfo{volume}{113}}, \bibinfo{pages}{107204}}
  (\href{http://dx.doi.org/10.1103/PhysRevLett.113.107204}{\bibinfo{year}{2014}}).

\bibitem{bera15}
\bibinfo{author}{S.~Bera}, \bibinfo{author}{H.~Schomerus},
  \bibinfo{author}{F.~Heidrich-Meisner}, and \bibinfo{author}{J.~H. Bardarson},
  \bibinfo{title}{Many-body localization characterized from a one-particle
  perspective},
  \bibinfo{journal}{\href{http://dx.doi.org/10.1103/PhysRevLett.115.046603}{Phys.
  Rev. Lett.}} \href{http://dx.doi.org/10.1103/PhysRevLett.115.046603}{{\bf
  \bibinfo{volume}{115}}, \bibinfo{pages}{046603}}
  (\href{http://dx.doi.org/10.1103/PhysRevLett.115.046603}{\bibinfo{year}{2015}}).

\bibitem{lev15}
\bibinfo{author}{Y.~Bar~Lev}, \bibinfo{author}{G.~Cohen}, and
  \bibinfo{author}{D.~R. Reichman}, \bibinfo{title}{Absence of diffusion in an
  interacting system of spinless fermions on a one-dimensional disordered
  lattice},
  \bibinfo{journal}{\href{http://dx.doi.org/10.1103/PhysRevLett.114.100601}{Phys.
  Rev. Lett.}} \href{http://dx.doi.org/10.1103/PhysRevLett.114.100601}{{\bf
  \bibinfo{volume}{114}}, \bibinfo{pages}{100601}}
  (\href{http://dx.doi.org/10.1103/PhysRevLett.114.100601}{\bibinfo{year}{2015}}).

\bibitem{serbyn2017}
\bibinfo{author}{M.~Serbyn}, \bibinfo{author}{Z.~Papi\ifmmode~\acute{c}\else
  \'{c}\fi{}}, and \bibinfo{author}{D.~A. Abanin}, \bibinfo{title}{Thouless
  energy and multifractality across the many-body localization transition},
  \bibinfo{journal}{\href{http://dx.doi.org/10.1103/PhysRevB.96.104201}{Phys.
  Rev. B}} \href{http://dx.doi.org/10.1103/PhysRevB.96.104201}{{\bf
  \bibinfo{volume}{96}}, \bibinfo{pages}{104201}}
  (\href{http://dx.doi.org/10.1103/PhysRevB.96.104201}{\bibinfo{year}{2017}}).

\bibitem{sels2020}
\bibinfo{author}{D.~Sels} and \bibinfo{author}{A.~Polkovnikov},
  \bibinfo{title}{Dynamical obstruction to localization in a disordered spin
  chain},
  \bibinfo{journal}{\href{http://dx.doi.org/10.1103/PhysRevE.104.054105}{Phys.
  Rev. E}} \href{http://dx.doi.org/10.1103/PhysRevE.104.054105}{{\bf
  \bibinfo{volume}{104}}, \bibinfo{pages}{054105}}
  (\href{http://dx.doi.org/10.1103/PhysRevE.104.054105}{\bibinfo{year}{2021}}).

\bibitem{vidmar2021}
\bibinfo{author}{L.~Vidmar}, \bibinfo{author}{B.~Krajewski},
  \bibinfo{author}{J.~Bon\ifmmode~\check{c}\else \v{c}\fi{}a}, and
  \bibinfo{author}{M.~Mierzejewski}, \bibinfo{title}{Phenomenology of spectral
  functions in disordered spin chains at infinite temperature},
  \bibinfo{journal}{\href{http://dx.doi.org/10.1103/PhysRevLett.127.230603}{Phys.
  Rev. Lett.}} \href{http://dx.doi.org/10.1103/PhysRevLett.127.230603}{{\bf
  \bibinfo{volume}{127}}, \bibinfo{pages}{230603}}
  (\href{http://dx.doi.org/10.1103/PhysRevLett.127.230603}{\bibinfo{year}{2021}}).

\bibitem{luitz2016prl}
\bibinfo{author}{D.~J. Luitz} and \bibinfo{author}{Y.~Bar~Lev},
  \bibinfo{title}{Anomalous thermalization in ergodic systems},
  \bibinfo{journal}{\href{http://dx.doi.org/10.1103/PhysRevLett.117.170404}{Phys.
  Rev. Lett.}} \href{http://dx.doi.org/10.1103/PhysRevLett.117.170404}{{\bf
  \bibinfo{volume}{117}}, \bibinfo{pages}{170404}}
  (\href{http://dx.doi.org/10.1103/PhysRevLett.117.170404}{\bibinfo{year}{2016}}).

\bibitem{luitz116}
\bibinfo{author}{D.~J. Luitz} and \bibinfo{author}{Y.~Bar~Lev},
  \bibinfo{title}{The ergodic side of the many‐body localization transition},
  \bibinfo{journal}{\href{http://dx.doi.org/10.1002/andp.201600350}{Ann. Phys.
  (Berl.)}} \href{http://dx.doi.org/10.1002/andp.201600350}{{\bf
  \bibinfo{volume}{529}}, \bibinfo{pages}{1600350}}
  (\href{http://dx.doi.org/10.1002/andp.201600350}{\bibinfo{year}{2016}}).

\bibitem{znidaric16}
\bibinfo{author}{M.~{\v Znidari\v c}}, \bibinfo{author}{A.~Scardicchio}, and
  \bibinfo{author}{V.~K. Varma}, \bibinfo{title}{Diffusive and subdiffusive
  spin transport in the ergodic phase of a many-body localizable system},
  \bibinfo{journal}{\href{http://dx.doi.org/10.1103/PhysRevLett.117.040601}{Phys.
  Rev. Lett.}} \href{http://dx.doi.org/10.1103/PhysRevLett.117.040601}{{\bf
  \bibinfo{volume}{117}}, \bibinfo{pages}{040601}}
  (\href{http://dx.doi.org/10.1103/PhysRevLett.117.040601}{\bibinfo{year}{2016}}).

\bibitem{gopal17}
\bibinfo{author}{S.~Gopalakrishnan}, \bibinfo{author}{K.~R. Islam}, and
  \bibinfo{author}{M.~Knap}, \bibinfo{title}{Noise-induced subdiffusion in
  strongly localized quantum systems},
  \bibinfo{journal}{\href{http://dx.doi.org/10.1103/PhysRevLett.119.046601}{Phys.
  Rev. Lett.}} \href{http://dx.doi.org/10.1103/PhysRevLett.119.046601}{{\bf
  \bibinfo{volume}{119}}, \bibinfo{pages}{046601}}
  (\href{http://dx.doi.org/10.1103/PhysRevLett.119.046601}{\bibinfo{year}{2017}}).

\bibitem{kozarzewski18}
\bibinfo{author}{M.~Kozarzewski},
  \bibinfo{author}{P.~Prelov\ifmmode~\check{s}\else \v{s}\fi{}ek}, and
  \bibinfo{author}{M.~Mierzejewski}, \bibinfo{title}{Spin subdiffusion in the
  disordered {Hubbard} chain},
  \bibinfo{journal}{\href{http://dx.doi.org/10.1103/PhysRevLett.120.246602}{Phys.
  Rev. Lett.}} \href{http://dx.doi.org/10.1103/PhysRevLett.120.246602}{{\bf
  \bibinfo{volume}{120}}, \bibinfo{pages}{246602}}
  (\href{http://dx.doi.org/10.1103/PhysRevLett.120.246602}{\bibinfo{year}{2018}}).

\bibitem{prelovsek217}
\bibinfo{author}{P.~Prelov\ifmmode~\check{s}\else \v{s}\fi{}ek} and
  \bibinfo{author}{J.~Herbrych}, \bibinfo{title}{Self-consistent approach to
  many-body localization and subdiffusion},
  \bibinfo{journal}{\href{http://dx.doi.org/10.1103/PhysRevB.96.035130}{Phys.
  Rev. B}} \href{http://dx.doi.org/10.1103/PhysRevB.96.035130}{{\bf
  \bibinfo{volume}{96}}, \bibinfo{pages}{035130}}
  (\href{http://dx.doi.org/10.1103/PhysRevB.96.035130}{\bibinfo{year}{2017}}).

\bibitem{new_karrasch}
\bibinfo{author}{Y.~B. Lev}, \bibinfo{author}{D.~M. Kennes},
  \bibinfo{author}{C.~Kl{\"o}ckner}, \bibinfo{author}{D.~R. Reichman}, and
  \bibinfo{author}{C.~Karrasch}, \bibinfo{title}{{Transport in quasiperiodic
  interacting systems: From superdiffusion to subdiffusion}},
  \bibinfo{journal}{\href{http://dx.doi.org/10.1209/0295-5075/119/37003}{EPL
  (Europhysics Letters)}}
  \href{http://dx.doi.org/10.1209/0295-5075/119/37003}{{\bf
  \bibinfo{volume}{119}}, \bibinfo{pages}{37003}}
  (\href{http://dx.doi.org/10.1209/0295-5075/119/37003}{\bibinfo{year}{2017}}).

\bibitem{prelovsek2018a}
\bibinfo{author}{P.~Prelov\ifmmode~\check{s}\else \v{s}\fi{}ek},
  \bibinfo{author}{J.~Bon\ifmmode~\check{c}\else \v{c}\fi{}a}, and
  \bibinfo{author}{M.~Mierzejewski}, \bibinfo{title}{Transient and persistent
  particle subdiffusion in a disordered chain coupled to bosons},
  \bibinfo{journal}{\href{http://dx.doi.org/10.1103/PhysRevB.98.125119}{Phys.
  Rev. B}} \href{http://dx.doi.org/10.1103/PhysRevB.98.125119}{{\bf
  \bibinfo{volume}{98}}, \bibinfo{pages}{125119}}
  (\href{http://dx.doi.org/10.1103/PhysRevB.98.125119}{\bibinfo{year}{2018}}).

\bibitem{pal10}
\bibinfo{author}{A.~Pal} and \bibinfo{author}{D.~A. Huse},
  \bibinfo{title}{Many-body localization phase transition},
  \bibinfo{journal}{\href{http://dx.doi.org/10.1103/PhysRevB.82.174411}{Phys.
  Rev. B}} \href{http://dx.doi.org/10.1103/PhysRevB.82.174411}{{\bf
  \bibinfo{volume}{82}}, \bibinfo{pages}{174411}}
  (\href{http://dx.doi.org/10.1103/PhysRevB.82.174411}{\bibinfo{year}{2010}}).

\bibitem{monthus10}
\bibinfo{author}{C.~Monthus} and \bibinfo{author}{T.~Garel},
  \bibinfo{title}{Many-body localization transition in a lattice model of
  interacting fermions: Statistics of renormalized hoppings in configuration
  space},
  \bibinfo{journal}{\href{http://dx.doi.org/10.1103/PhysRevB.81.134202}{Phys.
  Rev. B}} \href{http://dx.doi.org/10.1103/PhysRevB.81.134202}{{\bf
  \bibinfo{volume}{81}}, \bibinfo{pages}{134202}}
  (\href{http://dx.doi.org/10.1103/PhysRevB.81.134202}{\bibinfo{year}{2010}}).

\bibitem{khatami_rigol_12}
\bibinfo{author}{E.~Khatami}, \bibinfo{author}{M.~Rigol},
  \bibinfo{author}{A.~Rela\~no}, and \bibinfo{author}{A.~M.
  Garc\'{\i}a-Garc\'{\i}a}, \bibinfo{title}{{Quantum quenches in disordered
  systems: Approach to thermal equilibrium without a typical relaxation time}},
  \bibinfo{journal}{\href{http://dx.doi.org/10.1103/PhysRevE.85.050102}{Phys.
  Rev. E}} \href{http://dx.doi.org/10.1103/PhysRevE.85.050102}{{\bf
  \bibinfo{volume}{85}}, \bibinfo{pages}{050102}}
  (\href{http://dx.doi.org/10.1103/PhysRevE.85.050102}{\bibinfo{year}{2012}}).

\bibitem{bardarson12}
\bibinfo{author}{J.~H. Bardarson}, \bibinfo{author}{F.~Pollmann}, and
  \bibinfo{author}{J.~E. Moore}, \bibinfo{title}{Unbounded growth of
  entanglement in models of many-body localization},
  \bibinfo{journal}{\href{http://dx.doi.org/10.1103/PhysRevLett.109.017202}{Phys.
  Rev. Lett.}} \href{http://dx.doi.org/10.1103/PhysRevLett.109.017202}{{\bf
  \bibinfo{volume}{109}}, \bibinfo{pages}{017202}}
  (\href{http://dx.doi.org/10.1103/PhysRevLett.109.017202}{\bibinfo{year}{2012}}).

\bibitem{deluca13}
\bibinfo{author}{A.~De~Luca} and \bibinfo{author}{A.~Scardicchio},
  \bibinfo{title}{Ergodicity breaking in a model showing many-body
  localization},
  \bibinfo{journal}{\href{http://dx.doi.org/10.1209/0295-5075/101/37003}{EPL
  (Europhysics Letters)}}
  \href{http://dx.doi.org/10.1209/0295-5075/101/37003}{{\bf
  \bibinfo{volume}{101}}, \bibinfo{pages}{37003}}
  (\href{http://dx.doi.org/10.1209/0295-5075/101/37003}{\bibinfo{year}{2013}}).

\bibitem{luitz15}
\bibinfo{author}{D.~J. Luitz}, \bibinfo{author}{N.~Laflorencie}, and
  \bibinfo{author}{F.~Alet}, \bibinfo{title}{Many-body localization edge in the
  random-field {Heisenberg} chain},
  \bibinfo{journal}{\href{http://dx.doi.org/10.1103/PhysRevB.91.081103}{Phys.
  Rev. B}} \href{http://dx.doi.org/10.1103/PhysRevB.91.081103}{{\bf
  \bibinfo{volume}{91}}, \bibinfo{pages}{081103}}
  (\href{http://dx.doi.org/10.1103/PhysRevB.91.081103}{\bibinfo{year}{2015}}).

\bibitem{torres15}
\bibinfo{author}{E.~J. Torres-Herrera} and \bibinfo{author}{L.~F. Santos},
  \bibinfo{title}{Dynamics at the many-body localization transition},
  \bibinfo{journal}{\href{http://dx.doi.org/10.1103/PhysRevB.92.014208}{Phys.
  Rev. B}} \href{http://dx.doi.org/10.1103/PhysRevB.92.014208}{{\bf
  \bibinfo{volume}{92}}, \bibinfo{pages}{014208}}
  (\href{http://dx.doi.org/10.1103/PhysRevB.92.014208}{\bibinfo{year}{2015}}).

\bibitem{serbyn15}
\bibinfo{author}{M.~Serbyn}, \bibinfo{author}{Z.~Papi{\'{c}}}, and
  \bibinfo{author}{D.~A. Abanin}, \bibinfo{title}{Criterion for many-body
  localization-delocalization phase transition},
  \bibinfo{journal}{\href{http://dx.doi.org/10.1103/PhysRevX.5.041047}{Phys.
  Rev. X}} \href{http://dx.doi.org/10.1103/PhysRevX.5.041047}{{\bf
  \bibinfo{volume}{5}}, \bibinfo{pages}{041047}}
  (\href{http://dx.doi.org/10.1103/PhysRevX.5.041047}{\bibinfo{year}{2015}}).

\bibitem{devakul_singh_15}
\bibinfo{author}{T.~Devakul} and \bibinfo{author}{R.~R.~P. Singh},
  \bibinfo{title}{Early breakdown of area-law entanglement at the many-body
  delocalization transition},
  \bibinfo{journal}{\href{http://dx.doi.org/10.1103/PhysRevLett.115.187201}{Phys.
  Rev. Lett.}} \href{http://dx.doi.org/10.1103/PhysRevLett.115.187201}{{\bf
  \bibinfo{volume}{115}}, \bibinfo{pages}{187201}}
  (\href{http://dx.doi.org/10.1103/PhysRevLett.115.187201}{\bibinfo{year}{2015}}).

\bibitem{serbyn_moore_16}
\bibinfo{author}{M.~Serbyn} and \bibinfo{author}{J.~E. Moore},
  \bibinfo{title}{Spectral statistics across the many-body localization
  transition},
  \bibinfo{journal}{\href{http://dx.doi.org/10.1103/PhysRevB.93.041424}{Phys.
  Rev. B}} \href{http://dx.doi.org/10.1103/PhysRevB.93.041424}{{\bf
  \bibinfo{volume}{93}}, \bibinfo{pages}{041424}}
  (\href{http://dx.doi.org/10.1103/PhysRevB.93.041424}{\bibinfo{year}{2016}}).

\bibitem{bertrand_garcia_16}
\bibinfo{author}{C.~L. Bertrand} and \bibinfo{author}{A.~M.
  Garc\'{\i}a-Garc\'{\i}a}, \bibinfo{title}{{Anomalous Thouless energy and
  critical statistics on the metallic side of the many-body localization
  transition}},
  \bibinfo{journal}{\href{http://dx.doi.org/10.1103/PhysRevB.94.144201}{Phys.
  Rev. B}} \href{http://dx.doi.org/10.1103/PhysRevB.94.144201}{{\bf
  \bibinfo{volume}{94}}, \bibinfo{pages}{144201}}
  (\href{http://dx.doi.org/10.1103/PhysRevB.94.144201}{\bibinfo{year}{2016}}).

\bibitem{mierzejewski2016}
\bibinfo{author}{M.~Mierzejewski}, \bibinfo{author}{J.~Herbrych}, and
  \bibinfo{author}{P.~Prelov\ifmmode~\check{s}\else \v{s}\fi{}ek},
  \bibinfo{title}{Universal dynamics of density correlations at the transition
  to the many-body localized state},
  \bibinfo{journal}{\href{http://dx.doi.org/10.1103/PhysRevB.94.224207}{Phys.
  Rev. B}} \href{http://dx.doi.org/10.1103/PhysRevB.94.224207}{{\bf
  \bibinfo{volume}{94}}, \bibinfo{pages}{224207}}
  (\href{http://dx.doi.org/10.1103/PhysRevB.94.224207}{\bibinfo{year}{2016}}).

\bibitem{steinigeweg16}
\bibinfo{author}{R.~Steinigeweg}, \bibinfo{author}{J.~Herbrych},
  \bibinfo{author}{F.~Pollmann}, and \bibinfo{author}{W.~Brenig},
  \bibinfo{title}{Typicality approach to the optical conductivity in thermal
  and many-body localized phases},
  \bibinfo{journal}{\href{http://dx.doi.org/10.1103/PhysRevB.94.180401}{Phys.
  Rev. B}} \href{http://dx.doi.org/10.1103/PhysRevB.94.180401}{{\bf
  \bibinfo{volume}{94}}, \bibinfo{pages}{180401}}
  (\href{http://dx.doi.org/10.1103/PhysRevB.94.180401}{\bibinfo{year}{2016}}).

\bibitem{filippone_brouwer_16}
\bibinfo{author}{M.~Filippone}, \bibinfo{author}{P.~W. Brouwer},
  \bibinfo{author}{J.~Eisert}, and \bibinfo{author}{F.~von Oppen},
  \bibinfo{title}{Drude weight fluctuations in many-body localized systems},
  \bibinfo{journal}{\href{http://dx.doi.org/10.1103/PhysRevB.94.201112}{Phys.
  Rev. B}} \href{http://dx.doi.org/10.1103/PhysRevB.94.201112}{{\bf
  \bibinfo{volume}{94}}, \bibinfo{pages}{201112}}
  (\href{http://dx.doi.org/10.1103/PhysRevB.94.201112}{\bibinfo{year}{2016}}).

\bibitem{khemani_lim_17}
\bibinfo{author}{V.~Khemani}, \bibinfo{author}{S.~P. Lim},
  \bibinfo{author}{D.~N. Sheng}, and \bibinfo{author}{D.~A. Huse},
  \bibinfo{title}{Critical properties of the many-body localization
  transition},
  \bibinfo{journal}{\href{http://dx.doi.org/10.1103/PhysRevX.7.021013}{Phys.
  Rev. X}} \href{http://dx.doi.org/10.1103/PhysRevX.7.021013}{{\bf
  \bibinfo{volume}{7}}, \bibinfo{pages}{021013}}
  (\href{http://dx.doi.org/10.1103/PhysRevX.7.021013}{\bibinfo{year}{2017}}).

\bibitem{khemani_sheng_17}
\bibinfo{author}{V.~Khemani}, \bibinfo{author}{D.~N. Sheng}, and
  \bibinfo{author}{D.~A. Huse}, \bibinfo{title}{Two universality classes for
  the many-body localization transition},
  \bibinfo{journal}{\href{http://dx.doi.org/10.1103/PhysRevLett.119.075702}{Phys.
  Rev. Lett.}} \href{http://dx.doi.org/10.1103/PhysRevLett.119.075702}{{\bf
  \bibinfo{volume}{119}}, \bibinfo{pages}{075702}}
  (\href{http://dx.doi.org/10.1103/PhysRevLett.119.075702}{\bibinfo{year}{2017}}).

\bibitem{detomasi_bera_17}
\bibinfo{author}{G.~De~Tomasi}, \bibinfo{author}{S.~Bera},
  \bibinfo{author}{J.~H. Bardarson}, and \bibinfo{author}{F.~Pollmann},
  \bibinfo{title}{Quantum mutual information as a probe for many-body
  localization},
  \bibinfo{journal}{\href{http://dx.doi.org/10.1103/PhysRevLett.118.016804}{Phys.
  Rev. Lett.}} \href{http://dx.doi.org/10.1103/PhysRevLett.118.016804}{{\bf
  \bibinfo{volume}{118}}, \bibinfo{pages}{016804}}
  (\href{http://dx.doi.org/10.1103/PhysRevLett.118.016804}{\bibinfo{year}{2017}}).

\bibitem{pietracaprina_parisi_17}
\bibinfo{author}{F.~Pietracaprina}, \bibinfo{author}{G.~Parisi},
  \bibinfo{author}{A.~Mariano}, \bibinfo{author}{S.~Pascazio}, and
  \bibinfo{author}{A.~Scardicchio}, \bibinfo{title}{Entanglement critical
  length at the many-body localization transition},
  \bibinfo{journal}{\href{http://dx.doi.org/10.1088/1742-5468/aa9338}{J. Stat.
  Mech.}} \href{http://dx.doi.org/10.1088/1742-5468/aa9338}{{\bf
  \bibinfo{volume}{{\rm (2017)}}}, \bibinfo{pages}{113102}}.

\bibitem{gray_bose_18}
\bibinfo{author}{J.~Gray}, \bibinfo{author}{S.~Bose}, and
  \bibinfo{author}{A.~Bayat}, \bibinfo{title}{{Many-body localization
  transition: Schmidt gap, entanglement length, and scaling}},
  \bibinfo{journal}{\href{http://dx.doi.org/10.1103/PhysRevB.97.201105}{Phys.
  Rev. B}} \href{http://dx.doi.org/10.1103/PhysRevB.97.201105}{{\bf
  \bibinfo{volume}{97}}, \bibinfo{pages}{201105}}
  (\href{http://dx.doi.org/10.1103/PhysRevB.97.201105}{\bibinfo{year}{2018}}).

\bibitem{sierant_zakrzewski_19}
\bibinfo{author}{P.~Sierant} and \bibinfo{author}{J.~Zakrzewski},
  \bibinfo{title}{Level statistics across the many-body localization
  transition},
  \bibinfo{journal}{\href{http://dx.doi.org/10.1103/PhysRevB.99.104205}{Phys.
  Rev. B}} \href{http://dx.doi.org/10.1103/PhysRevB.99.104205}{{\bf
  \bibinfo{volume}{99}}, \bibinfo{pages}{104205}}
  (\href{http://dx.doi.org/10.1103/PhysRevB.99.104205}{\bibinfo{year}{2019}}).

\bibitem{schiulaz_torresherrera_19}
\bibinfo{author}{M.~Schiulaz}, \bibinfo{author}{E.~J. Torres-Herrera}, and
  \bibinfo{author}{L.~F. Santos}, \bibinfo{title}{Thouless and relaxation time
  scales in many-body quantum systems},
  \bibinfo{journal}{\href{http://dx.doi.org/10.1103/PhysRevB.99.174313}{Phys.
  Rev. B}} \href{http://dx.doi.org/10.1103/PhysRevB.99.174313}{{\bf
  \bibinfo{volume}{99}}, \bibinfo{pages}{174313}}
  (\href{http://dx.doi.org/10.1103/PhysRevB.99.174313}{\bibinfo{year}{2019}}).

\bibitem{mace_alet_19}
\bibinfo{author}{N.~Mac\'e}, \bibinfo{author}{F.~Alet}, and
  \bibinfo{author}{N.~Laflorencie}, \bibinfo{title}{Multifractal scalings
  across the many-body localization transition},
  \bibinfo{journal}{\href{http://dx.doi.org/10.1103/PhysRevLett.123.180601}{Phys.
  Rev. Lett.}} \href{http://dx.doi.org/10.1103/PhysRevLett.123.180601}{{\bf
  \bibinfo{volume}{123}}, \bibinfo{pages}{180601}}
  (\href{http://dx.doi.org/10.1103/PhysRevLett.123.180601}{\bibinfo{year}{2019}}).

\bibitem{roy_chalker_19}
\bibinfo{author}{S.~Roy}, \bibinfo{author}{J.~T. Chalker}, and
  \bibinfo{author}{D.~E. Logan}, \bibinfo{title}{{Percolation in Fock space as
  a proxy for many-body localization}},
  \bibinfo{journal}{\href{http://dx.doi.org/10.1103/PhysRevB.99.104206}{Phys.
  Rev. B}} \href{http://dx.doi.org/10.1103/PhysRevB.99.104206}{{\bf
  \bibinfo{volume}{99}}, \bibinfo{pages}{104206}}
  (\href{http://dx.doi.org/10.1103/PhysRevB.99.104206}{\bibinfo{year}{2019}}).

\bibitem{roy_logan_20}
\bibinfo{author}{S.~Roy} and \bibinfo{author}{D.~E. Logan},
  \bibinfo{title}{Fock-space correlations and the origins of many-body
  localization},
  \bibinfo{journal}{\href{http://dx.doi.org/10.1103/PhysRevB.101.134202}{Phys.
  Rev. B}} \href{http://dx.doi.org/10.1103/PhysRevB.101.134202}{{\bf
  \bibinfo{volume}{101}}, \bibinfo{pages}{134202}}
  (\href{http://dx.doi.org/10.1103/PhysRevB.101.134202}{\bibinfo{year}{2020}}).

\bibitem{tarzia_20}
\bibinfo{author}{M.~Tarzia}, \bibinfo{title}{{Many-body localization transition
  in Hilbert space}},
  \bibinfo{journal}{\href{http://dx.doi.org/10.1103/PhysRevB.102.014208}{Phys.
  Rev. B}} \href{http://dx.doi.org/10.1103/PhysRevB.102.014208}{{\bf
  \bibinfo{volume}{102}}, \bibinfo{pages}{014208}}
  (\href{http://dx.doi.org/10.1103/PhysRevB.102.014208}{\bibinfo{year}{2020}}).

\bibitem{suntajs_2020}
\bibinfo{author}{J.~\v{S}untajs}, \bibinfo{author}{J.~Bon\v{c}a},
  \bibinfo{author}{T.~Prosen}, and \bibinfo{author}{L.~Vidmar},
  \bibinfo{title}{Ergodicity breaking transition in finite disordered spin
  chains},
  \bibinfo{journal}{\href{http://dx.doi.org/10.1103/PhysRevB.102.064207}{Phys.
  Rev. B}} \href{http://dx.doi.org/10.1103/PhysRevB.102.064207}{{\bf
  \bibinfo{volume}{102}}, \bibinfo{pages}{064207}}
  (\href{http://dx.doi.org/10.1103/PhysRevB.102.064207}{\bibinfo{year}{2020}}).

\bibitem{suntajs_bonca_20a}
\bibinfo{author}{J.~\v{S}untajs}, \bibinfo{author}{J.~Bon\v{c}a},
  \bibinfo{author}{T.~Prosen}, and \bibinfo{author}{L.~Vidmar},
  \bibinfo{title}{Quantum chaos challenges many-body localization},
  \bibinfo{journal}{\href{http://dx.doi.org/10.1103/PhysRevE.102.062144}{Phys.
  Rev. E}} \href{http://dx.doi.org/10.1103/PhysRevE.102.062144}{{\bf
  \bibinfo{volume}{102}}, \bibinfo{pages}{062144}}
  (\href{http://dx.doi.org/10.1103/PhysRevE.102.062144}{\bibinfo{year}{2020}}).

\bibitem{kieferemmanouilidis_unanyan_20}
\bibinfo{author}{M.~Kiefer-Emmanouilidis}, \bibinfo{author}{R.~Unanyan},
  \bibinfo{author}{M.~Fleischhauer}, and \bibinfo{author}{J.~Sirker},
  \bibinfo{title}{Evidence for unbounded growth of the number entropy in
  many-body localized phases},
  \bibinfo{journal}{\href{http://dx.doi.org/10.1103/PhysRevLett.124.243601}{Phys.
  Rev. Lett.}} \href{http://dx.doi.org/10.1103/PhysRevLett.124.243601}{{\bf
  \bibinfo{volume}{124}}, \bibinfo{pages}{243601}}
  (\href{http://dx.doi.org/10.1103/PhysRevLett.124.243601}{\bibinfo{year}{2020}}).

\bibitem{kieferemmanouilidis_unanyan_21}
\bibinfo{author}{M.~Kiefer-Emmanouilidis}, \bibinfo{author}{R.~Unanyan},
  \bibinfo{author}{M.~Fleischhauer}, and \bibinfo{author}{J.~Sirker},
  \bibinfo{title}{Slow delocalization of particles in many-body localized
  phases},
  \bibinfo{journal}{\href{http://dx.doi.org/10.1103/PhysRevB.103.024203}{Phys.
  Rev. B}} \href{http://dx.doi.org/10.1103/PhysRevB.103.024203}{{\bf
  \bibinfo{volume}{103}}, \bibinfo{pages}{024203}}
  (\href{http://dx.doi.org/10.1103/PhysRevB.103.024203}{\bibinfo{year}{2021}}).

\bibitem{leblond_sels_21}
\bibinfo{author}{T.~LeBlond}, \bibinfo{author}{D.~Sels},
  \bibinfo{author}{A.~Polkovnikov}, and \bibinfo{author}{M.~Rigol},
  \bibinfo{title}{Universality in the onset of quantum chaos in many-body
  systems},
  \bibinfo{journal}{\href{http://dx.doi.org/10.1103/PhysRevB.104.L201117}{Phys.
  Rev. B}} \href{http://dx.doi.org/10.1103/PhysRevB.104.L201117}{{\bf
  \bibinfo{volume}{104}}, \bibinfo{pages}{L201117}}
  (\href{http://dx.doi.org/10.1103/PhysRevB.104.L201117}{\bibinfo{year}{2021}}).

\bibitem{Sels_dilute_2021}
\bibinfo{author}{D.~Sels} and \bibinfo{author}{A.~Polkovnikov},
  \bibinfo{title}{Thermalization of dilute impurities in one dimensional spin
  chains},
  \href{https://arxiv.org/abs/2105.09348}{\bibinfo{howpublished}{arXiv:2105.09348}}
  (\bibinfo{year}{2021}).

\bibitem{deroeck_huveneers_17}
\bibinfo{author}{W.~De~Roeck} and \bibinfo{author}{F.~Huveneers},
  \bibinfo{title}{Stability and instability towards delocalization in many-body
  localization systems},
  \bibinfo{journal}{\href{http://dx.doi.org/10.1103/PhysRevB.95.155129}{Phys.
  Rev. B}} \href{http://dx.doi.org/10.1103/PhysRevB.95.155129}{{\bf
  \bibinfo{volume}{95}}, \bibinfo{pages}{155129}}
  (\href{http://dx.doi.org/10.1103/PhysRevB.95.155129}{\bibinfo{year}{2017}}).

\bibitem{luitz_huveneers_17}
\bibinfo{author}{D.~J. Luitz}, \bibinfo{author}{F.~Huveneers}, and
  \bibinfo{author}{W.~De~Roeck}, \bibinfo{title}{How a small quantum bath can
  thermalize long localized chains},
  \bibinfo{journal}{\href{http://dx.doi.org/10.1103/PhysRevLett.119.150602}{Phys.
  Rev. Lett.}} \href{http://dx.doi.org/10.1103/PhysRevLett.119.150602}{{\bf
  \bibinfo{volume}{119}}, \bibinfo{pages}{150602}}
  (\href{http://dx.doi.org/10.1103/PhysRevLett.119.150602}{\bibinfo{year}{2017}}).

\bibitem{thiery_huveneers_18}
\bibinfo{author}{T.~Thiery}, \bibinfo{author}{F.~Huveneers},
  \bibinfo{author}{M.~M\"uller}, and \bibinfo{author}{W.~De~Roeck},
  \bibinfo{title}{Many-body delocalization as a quantum avalanche},
  \bibinfo{journal}{\href{http://dx.doi.org/10.1103/PhysRevLett.121.140601}{Phys.
  Rev. Lett.}} \href{http://dx.doi.org/10.1103/PhysRevLett.121.140601}{{\bf
  \bibinfo{volume}{121}}, \bibinfo{pages}{140601}}
  (\href{http://dx.doi.org/10.1103/PhysRevLett.121.140601}{\bibinfo{year}{2018}}).

\bibitem{crowley_chandran_20}
\bibinfo{author}{P.~J.~D. Crowley} and \bibinfo{author}{A.~Chandran},
  \bibinfo{title}{Avalanche induced coexisting localized and thermal regions in
  disordered chains},
  \bibinfo{journal}{\href{http://dx.doi.org/10.1103/PhysRevResearch.2.033262}{Phys.
  Rev. Research}}
  \href{http://dx.doi.org/10.1103/PhysRevResearch.2.033262}{{\bf
  \bibinfo{volume}{2}}, \bibinfo{pages}{033262}}
  (\href{http://dx.doi.org/10.1103/PhysRevResearch.2.033262}{\bibinfo{year}{2020}}).

\bibitem{Sels_2022}
\bibinfo{author}{D.~Sels}, \bibinfo{title}{Bath-induced delocalization in
  interacting disordered spin chains},
  \bibinfo{journal}{\href{http://dx.doi.org/10.1103/physrevb.106.l020202}{Phys.
  Rev. B}} \href{http://dx.doi.org/10.1103/physrevb.106.l020202}{{\bf
  \bibinfo{volume}{106}}}
  (\href{http://dx.doi.org/10.1103/physrevb.106.l020202}{\bibinfo{year}{2022}}).

\bibitem{Morningstar2022}
\bibinfo{author}{A.~Morningstar}, \bibinfo{author}{L.~Colmenarez},
  \bibinfo{author}{V.~Khemani}, \bibinfo{author}{D.~J. Luitz}, and
  \bibinfo{author}{D.~A. Huse}, \bibinfo{title}{Avalanches and many-body
  resonances in many-body localized systems},
  \bibinfo{journal}{\href{http://dx.doi.org/10.1103/PhysRevB.105.174205}{Phys.
  Rev. B}} \href{http://dx.doi.org/10.1103/PhysRevB.105.174205}{{\bf
  \bibinfo{volume}{105}}, \bibinfo{pages}{174205}}
  (\href{http://dx.doi.org/10.1103/PhysRevB.105.174205}{\bibinfo{year}{2022}}).

\bibitem{suntajs_vidmar_22}
\bibinfo{author}{J.~\ifmmode~\check{S}\else \v{S}\fi{}untajs} and
  \bibinfo{author}{L.~Vidmar}, \bibinfo{title}{{Ergodicity Breaking Transition
  in Zero Dimensions}},
  \bibinfo{journal}{\href{http://dx.doi.org/10.1103/PhysRevLett.129.060602}{Phys.
  Rev. Lett.}} \href{http://dx.doi.org/10.1103/PhysRevLett.129.060602}{{\bf
  \bibinfo{volume}{129}}, \bibinfo{pages}{060602}}
  (\href{http://dx.doi.org/10.1103/PhysRevLett.129.060602}{\bibinfo{year}{2022}}).

\bibitem{crowley_chandran_22b}
\bibinfo{author}{P.~J.~D. Crowley} and \bibinfo{author}{A.~Chandran},
  \bibinfo{title}{Mean field theory of failed thermalizing avalanches},
  \href{https://arxiv.org/abs/2204.09688}{\bibinfo{howpublished}{arXiv:2204.09688}}.

\bibitem{Panda2020}
\bibinfo{author}{R.~K. Panda}, \bibinfo{author}{A.~Scardicchio},
  \bibinfo{author}{M.~Schulz}, \bibinfo{author}{S.~R. Taylor}, and
  \bibinfo{author}{M.~{\v{Z}}nidari{\v{c}}}, \bibinfo{title}{Can we study the
  many-body localisation transition?},
  \bibinfo{journal}{\href{http://dx.doi.org/10.1209/0295-5075/128/67003}{{EPL}
  (Europhysics Letters)}}
  \href{http://dx.doi.org/10.1209/0295-5075/128/67003}{{\bf
  \bibinfo{volume}{128}}, \bibinfo{pages}{67003}}
  (\href{http://dx.doi.org/10.1209/0295-5075/128/67003}{\bibinfo{year}{2020}}).

\bibitem{Sierant2020}
\bibinfo{author}{P.~Sierant}, \bibinfo{author}{D.~Delande}, and
  \bibinfo{author}{J.~Zakrzewski}, \bibinfo{title}{Thouless time analysis of
  anderson and many-body localization transitions},
  \bibinfo{journal}{\href{http://dx.doi.org/10.1103/PhysRevLett.124.186601}{Phys.
  Rev. Lett.}} \href{http://dx.doi.org/10.1103/PhysRevLett.124.186601}{{\bf
  \bibinfo{volume}{124}}, \bibinfo{pages}{186601}}
  (\href{http://dx.doi.org/10.1103/PhysRevLett.124.186601}{\bibinfo{year}{2020}}).

\bibitem{sierant_lewenstein_20}
\bibinfo{author}{P.~Sierant}, \bibinfo{author}{M.~Lewenstein}, and
  \bibinfo{author}{J.~Zakrzewski}, \bibinfo{title}{Polynomially filtered exact
  diagonalization approach to many-body localization},
  \bibinfo{journal}{\href{http://dx.doi.org/10.1103/PhysRevLett.125.156601}{Phys.
  Rev. Lett.}} \href{http://dx.doi.org/10.1103/PhysRevLett.125.156601}{{\bf
  \bibinfo{volume}{125}}, \bibinfo{pages}{156601}}
  (\href{http://dx.doi.org/10.1103/PhysRevLett.125.156601}{\bibinfo{year}{2020}}).

\bibitem{abanin_bardarson_21}
\bibinfo{author}{D.~Abanin}, \bibinfo{author}{J.~Bardarson},
  \bibinfo{author}{G.~{De Tomasi}}, \bibinfo{author}{S.~Gopalakrishnan},
  \bibinfo{author}{V.~Khemani}, \bibinfo{author}{S.~Parameswaran},
  \bibinfo{author}{F.~Pollmann}, \bibinfo{author}{A.~Potter},
  \bibinfo{author}{M.~Serbyn}, and \bibinfo{author}{R.~Vasseur},
  \bibinfo{title}{{Distinguishing localization from chaos: Challenges in
  finite-size systems}},
  \bibinfo{journal}{\href{http://dx.doi.org/https://doi.org/10.1016/j.aop.2021.168415}{Ann.
  Phys.}}
  \href{http://dx.doi.org/https://doi.org/10.1016/j.aop.2021.168415}{{\bf
  \bibinfo{volume}{427}}, \bibinfo{pages}{168415}}
  (\href{http://dx.doi.org/https://doi.org/10.1016/j.aop.2021.168415}{\bibinfo{year}{2021}}).

\bibitem{corps_molina_21}
\bibinfo{author}{{\'A}.~L. Corps}, \bibinfo{author}{R.~A. Molina}, and
  \bibinfo{author}{A.~Rela{\~n}o}, \bibinfo{title}{{Signatures of a critical
  point in the many-body localization transition}},
  \bibinfo{journal}{\href{http://dx.doi.org/10.21468/SciPostPhys.10.5.107}{SciPost
  Phys.}} \href{http://dx.doi.org/10.21468/SciPostPhys.10.5.107}{{\bf
  \bibinfo{volume}{10}}, \bibinfo{pages}{107}}
  (\href{http://dx.doi.org/10.21468/SciPostPhys.10.5.107}{\bibinfo{year}{2021}}).

\bibitem{prakash_pixley_21}
\bibinfo{author}{A.~Prakash}, \bibinfo{author}{J.~H. Pixley}, and
  \bibinfo{author}{M.~Kulkarni}, \bibinfo{title}{Universal spectral form factor
  for many-body localization},
  \bibinfo{journal}{\href{http://dx.doi.org/10.1103/PhysRevResearch.3.L012019}{Phys.
  Rev. Research}}
  \href{http://dx.doi.org/10.1103/PhysRevResearch.3.L012019}{{\bf
  \bibinfo{volume}{3}}, \bibinfo{pages}{L012019}}
  (\href{http://dx.doi.org/10.1103/PhysRevResearch.3.L012019}{\bibinfo{year}{2021}}).

\bibitem{schliemann_costa_21}
\bibinfo{author}{J.~Schliemann}, \bibinfo{author}{J.~V.~I. Costa},
  \bibinfo{author}{P.~Wenk}, and \bibinfo{author}{J.~C. Egues},
  \bibinfo{title}{{Many-body localization: Transitions in spin models}},
  \bibinfo{journal}{\href{http://dx.doi.org/10.1103/PhysRevB.103.174203}{Phys.
  Rev. B}} \href{http://dx.doi.org/10.1103/PhysRevB.103.174203}{{\bf
  \bibinfo{volume}{103}}, \bibinfo{pages}{174203}}
  (\href{http://dx.doi.org/10.1103/PhysRevB.103.174203}{\bibinfo{year}{2021}}).

\bibitem{hopjan_orso_21}
\bibinfo{author}{M.~Hopjan}, \bibinfo{author}{G.~Orso}, and
  \bibinfo{author}{F.~Heidrich-Meisner}, \bibinfo{title}{Detecting
  delocalization-localization transitions from full density distributions},
  \bibinfo{journal}{\href{http://dx.doi.org/10.1103/PhysRevB.104.235112}{Phys.
  Rev. B}} \href{http://dx.doi.org/10.1103/PhysRevB.104.235112}{{\bf
  \bibinfo{volume}{104}}, \bibinfo{pages}{235112}}
  (\href{http://dx.doi.org/10.1103/PhysRevB.104.235112}{\bibinfo{year}{2021}}).

\bibitem{solorzano_santos_21}
\bibinfo{author}{A.~Sol\'orzano}, \bibinfo{author}{L.~F. Santos}, and
  \bibinfo{author}{E.~J. Torres-Herrera}, \bibinfo{title}{Multifractality and
  self-averaging at the many-body localization transition},
  \bibinfo{journal}{\href{http://dx.doi.org/10.1103/PhysRevResearch.3.L032030}{Phys.
  Rev. Research}}
  \href{http://dx.doi.org/10.1103/PhysRevResearch.3.L032030}{{\bf
  \bibinfo{volume}{3}}, \bibinfo{pages}{L032030}}
  (\href{http://dx.doi.org/10.1103/PhysRevResearch.3.L032030}{\bibinfo{year}{2021}}).

\bibitem{detomasi_khaymovich_21}
\bibinfo{author}{G.~De~Tomasi}, \bibinfo{author}{I.~M. Khaymovich},
  \bibinfo{author}{F.~Pollmann}, and \bibinfo{author}{S.~Warzel},
  \bibinfo{title}{{Rare thermal bubbles at the many-body localization
  transition from the Fock space point of view}},
  \bibinfo{journal}{\href{http://dx.doi.org/10.1103/PhysRevB.104.024202}{Phys.
  Rev. B}} \href{http://dx.doi.org/10.1103/PhysRevB.104.024202}{{\bf
  \bibinfo{volume}{104}}, \bibinfo{pages}{024202}}
  (\href{http://dx.doi.org/10.1103/PhysRevB.104.024202}{\bibinfo{year}{2021}}).

\bibitem{crowley_chandran_22}
\bibinfo{author}{P.~J.~D. Crowley} and \bibinfo{author}{A.~Chandran},
  \bibinfo{title}{{A constructive theory of the numerically accessible
  many-body localized to thermal crossover}},
  \bibinfo{journal}{\href{http://dx.doi.org/10.21468/SciPostPhys.12.6.201}{SciPost
  Phys.}} \href{http://dx.doi.org/10.21468/SciPostPhys.12.6.201}{{\bf
  \bibinfo{volume}{12}}, \bibinfo{pages}{201}}
  (\href{http://dx.doi.org/10.21468/SciPostPhys.12.6.201}{\bibinfo{year}{2022}}).

\bibitem{ghosh_znidaric_22}
\bibinfo{author}{R.~Ghosh} and \bibinfo{author}{M.~\ifmmode \check{Z}\else
  \v{Z}\fi{}nidari\ifmmode~\check{c}\else \v{c}\fi{}},
  \bibinfo{title}{Resonance-induced growth of number entropy in strongly
  disordered systems},
  \bibinfo{journal}{\href{http://dx.doi.org/10.1103/PhysRevB.105.144203}{Phys.
  Rev. B}} \href{http://dx.doi.org/10.1103/PhysRevB.105.144203}{{\bf
  \bibinfo{volume}{105}}, \bibinfo{pages}{144203}}
  (\href{http://dx.doi.org/10.1103/PhysRevB.105.144203}{\bibinfo{year}{2022}}).

\bibitem{bolther_kehrein_22}
\bibinfo{author}{N.~B\"olter} and \bibinfo{author}{S.~Kehrein},
  \bibinfo{title}{{Scrambling and many-body localization in the XXZ chain}},
  \bibinfo{journal}{\href{http://dx.doi.org/10.1103/PhysRevB.105.104202}{Phys.
  Rev. B}} \href{http://dx.doi.org/10.1103/PhysRevB.105.104202}{{\bf
  \bibinfo{volume}{105}}, \bibinfo{pages}{104202}}
  (\href{http://dx.doi.org/10.1103/PhysRevB.105.104202}{\bibinfo{year}{2022}}).

\bibitem{yintai_yufeng_22}
\bibinfo{author}{Y.~Zhang} and \bibinfo{author}{Y.~Liang},
  \bibinfo{title}{Optimizing randomized potentials for inhibiting
  thermalization in one-dimensional systems},
  \bibinfo{journal}{\href{http://dx.doi.org/10.1103/PhysRevResearch.4.023091}{Phys.
  Rev. Research}}
  \href{http://dx.doi.org/10.1103/PhysRevResearch.4.023091}{{\bf
  \bibinfo{volume}{4}}, \bibinfo{pages}{023091}}
  (\href{http://dx.doi.org/10.1103/PhysRevResearch.4.023091}{\bibinfo{year}{2022}}).

\bibitem{sierant2022}
\bibinfo{author}{P.~Sierant} and \bibinfo{author}{J.~Zakrzewski},
  \bibinfo{title}{Challenges to observation of many-body localization},
  \bibinfo{journal}{\href{http://dx.doi.org/10.1103/PhysRevB.105.224203}{Phys.
  Rev. B}} \href{http://dx.doi.org/10.1103/PhysRevB.105.224203}{{\bf
  \bibinfo{volume}{105}}, \bibinfo{pages}{224203}}
  (\href{http://dx.doi.org/10.1103/PhysRevB.105.224203}{\bibinfo{year}{2022}}).

\bibitem{sutradhar_ghosh_22}
\bibinfo{author}{J.~Sutradhar}, \bibinfo{author}{S.~Ghosh},
  \bibinfo{author}{S.~Roy}, \bibinfo{author}{D.~E. Logan},
  \bibinfo{author}{S.~Mukerjee}, and \bibinfo{author}{S.~Banerjee},
  \bibinfo{title}{{Scaling of the Fock-space propagator and multifractality
  across the many-body localization transition}},
  \bibinfo{journal}{\href{http://dx.doi.org/10.1103/PhysRevB.106.054203}{Phys.
  Rev. B}} \href{http://dx.doi.org/10.1103/PhysRevB.106.054203}{{\bf
  \bibinfo{volume}{106}}, \bibinfo{pages}{054203}}
  (\href{http://dx.doi.org/10.1103/PhysRevB.106.054203}{\bibinfo{year}{2022}}).

\bibitem{trigueros_cheng_22}
\bibinfo{author}{F.~B. Trigueros} and \bibinfo{author}{C.-J. Lin},
  \bibinfo{title}{{Krylov complexity of many-body localization: Operator
  localization in Krylov basis}},
  \bibinfo{journal}{\href{http://dx.doi.org/10.21468/SciPostPhys.13.2.037}{SciPost
  Phys.}} \href{http://dx.doi.org/10.21468/SciPostPhys.13.2.037}{{\bf
  \bibinfo{volume}{13}}, \bibinfo{pages}{037}}
  (\href{http://dx.doi.org/10.21468/SciPostPhys.13.2.037}{\bibinfo{year}{2022}}).

\bibitem{shi_khemani_22}
\bibinfo{author}{D.~Z. Shi}, \bibinfo{author}{V.~Khemani},
  \bibinfo{author}{R.~Vasseur}, and \bibinfo{author}{S.~Gopalakrishnan},
  \bibinfo{title}{Many body localization transition with correlated disorder},
  \href{https://arxiv.org/abs/2204.06017}{\bibinfo{howpublished}{arXiv:2204.06017}}.

\bibitem{deutsch_91}
\bibinfo{author}{J.~M. Deutsch}, \bibinfo{title}{Quantum statistical mechanics
  in a closed system},
  \bibinfo{journal}{\href{http://dx.doi.org/10.1103/PhysRevA.43.2046}{Phys.
  Rev. A}} \href{http://dx.doi.org/10.1103/PhysRevA.43.2046}{{\bf
  \bibinfo{volume}{43}}, \bibinfo{pages}{2046}}
  (\href{http://dx.doi.org/10.1103/PhysRevA.43.2046}{\bibinfo{year}{1991}}).

\bibitem{srednicki_94}
\bibinfo{author}{M.~Srednicki}, \bibinfo{title}{Chaos and quantum
  thermalization},
  \bibinfo{journal}{\href{http://dx.doi.org/10.1103/PhysRevE.50.888}{Phys. Rev.
  E}} \href{http://dx.doi.org/10.1103/PhysRevE.50.888}{{\bf
  \bibinfo{volume}{50}}, \bibinfo{pages}{888}}
  (\href{http://dx.doi.org/10.1103/PhysRevE.50.888}{\bibinfo{year}{1994}}).

\bibitem{rigol_dunjko_08}
\bibinfo{author}{M.~Rigol}, \bibinfo{author}{V.~Dunjko}, and
  \bibinfo{author}{M.~Olshanii}, \bibinfo{title}{Thermalization and its
  mechanism for generic isolated quantum systems},
  \bibinfo{journal}{\href{http://dx.doi.org/10.1038/nature06838}{Nature
  (London)}} \href{http://dx.doi.org/10.1038/nature06838}{{\bf
  \bibinfo{volume}{452}}, \bibinfo{pages}{854}}
  (\href{http://dx.doi.org/10.1038/nature06838}{\bibinfo{year}{2008}}).

\bibitem{dalessio_kafri_16}
\bibinfo{author}{L.~D'Alessio}, \bibinfo{author}{Y.~Kafri},
  \bibinfo{author}{A.~Polkovnikov}, and \bibinfo{author}{M.~Rigol},
  \bibinfo{title}{From quantum chaos and eigenstate thermalization to
  statistical mechanics and thermodynamics},
  \bibinfo{journal}{\href{http://dx.doi.org/10.1080/00018732.2016.1198134}{Adv.
  Phys.}} \href{http://dx.doi.org/10.1080/00018732.2016.1198134}{{\bf
  \bibinfo{volume}{65}}, \bibinfo{pages}{239}}
  (\href{http://dx.doi.org/10.1080/00018732.2016.1198134}{\bibinfo{year}{2016}}).

\bibitem{Note1}
\bibinfo{note}{In contrast, in translationally invariant integrable models the
  products of local charges are typically nonlocal thus, in the thermodynamic
  limit, they have no projection on local observables.}

\bibitem{suppmat}
\bibinfo{note}{See {Supplemental Material} for reorthogonalization of LIOMs in
  the subspace with fixed number of fermions, absence of projection of
  $H_\Delta$ on LIOMs, derivation of Eq.~(\ref{def_Hperpnorm}) and the lower
  bound on the norm of the true perturbation, the details about the nearest
  level spacing analysis and the fluctuations of the diagonal matrix elements.}

\bibitem{atas2013}
\bibinfo{author}{Y.~Y. Atas}, \bibinfo{author}{E.~Bogomolny},
  \bibinfo{author}{O.~Giraud}, and \bibinfo{author}{G.~Roux},
  \bibinfo{title}{Distribution of the ratio of consecutive level spacings in
  random matrix ensembles},
  \bibinfo{journal}{\href{http://dx.doi.org/10.1103/PhysRevLett.110.084101}{Phys.
  Rev. Lett.}} \href{http://dx.doi.org/10.1103/PhysRevLett.110.084101}{{\bf
  \bibinfo{volume}{110}}, \bibinfo{pages}{084101}}
  (\href{http://dx.doi.org/10.1103/PhysRevLett.110.084101}{\bibinfo{year}{2013}}).

\bibitem{fremling2022}
\bibinfo{author}{M.~Fremling}, \bibinfo{title}{Exact gap-ratio results for
  mixed {W}igner surmises of up to 4 eigenvalues},
  \href{https://arxiv.org/abs/2202.01090}{\bibinfo{howpublished}{arXiv:2202.01090}}
  (\bibinfo{year}{2022}).

\bibitem{alet2022}
\bibinfo{author}{O.~Giraud}, \bibinfo{author}{N.~Mac\'e},
  \bibinfo{author}{E.~Vernier}, and \bibinfo{author}{F.~Alet},
  \bibinfo{title}{Probing symmetries of quantum many-body systems through gap
  ratio statistics},
  \bibinfo{journal}{\href{http://dx.doi.org/10.1103/PhysRevX.12.011006}{Phys.
  Rev. X}} \href{http://dx.doi.org/10.1103/PhysRevX.12.011006}{{\bf
  \bibinfo{volume}{12}}, \bibinfo{pages}{011006}}
  (\href{http://dx.doi.org/10.1103/PhysRevX.12.011006}{\bibinfo{year}{2022}}).

\bibitem{schreiber15}
\bibinfo{author}{M.~Schreiber}, \bibinfo{author}{S.~S. Hodgman},
  \bibinfo{author}{P.~Bordia}, \bibinfo{author}{H.~P. L{\"{u}}schen},
  \bibinfo{author}{M.~H. Fischer}, \bibinfo{author}{R.~Vosk},
  \bibinfo{author}{E.~Altman}, \bibinfo{author}{U.~Schneider}, and
  \bibinfo{author}{I.~Bloch}, \bibinfo{title}{{Observation of many-body
  localization of interacting fermions in a quasi-random optical lattice}},
  \bibinfo{journal}{\href{http://dx.doi.org/10.1126/science.aaa7432}{Science}}
  \href{http://dx.doi.org/10.1126/science.aaa7432}{{\bf \bibinfo{volume}{349}},
  \bibinfo{pages}{842}}
  (\href{http://dx.doi.org/10.1126/science.aaa7432}{\bibinfo{year}{2015}}).

\bibitem{Kim_strong2014}
\bibinfo{author}{H.~Kim}, \bibinfo{author}{T.~N. Ikeda}, and
  \bibinfo{author}{D.~A. Huse}, \bibinfo{title}{Testing whether all eigenstates
  obey the eigenstate thermalization hypothesis},
  \bibinfo{journal}{\href{http://dx.doi.org/10.1103/PhysRevE.90.052105}{Phys.
  Rev. E}} \href{http://dx.doi.org/10.1103/PhysRevE.90.052105}{{\bf
  \bibinfo{volume}{90}}, \bibinfo{pages}{052105}}
  (\href{http://dx.doi.org/10.1103/PhysRevE.90.052105}{\bibinfo{year}{2014}}).

\bibitem{Mondaini2016}
\bibinfo{author}{R.~Mondaini}, \bibinfo{author}{K.~R. Fratus},
  \bibinfo{author}{M.~Srednicki}, and \bibinfo{author}{M.~Rigol},
  \bibinfo{title}{{Eigenstate thermalization in the two-dimensional transverse
  field Ising model}},
  \bibinfo{journal}{\href{http://dx.doi.org/10.1103/PhysRevE.93.032104}{Phys.
  Rev. E}} \href{http://dx.doi.org/10.1103/PhysRevE.93.032104}{{\bf
  \bibinfo{volume}{93}}, \bibinfo{pages}{032104}}
  (\href{http://dx.doi.org/10.1103/PhysRevE.93.032104}{\bibinfo{year}{2016}}).

\bibitem{jansen_stolpp_19}
\bibinfo{author}{D.~Jansen}, \bibinfo{author}{J.~Stolpp},
  \bibinfo{author}{L.~Vidmar}, and \bibinfo{author}{F.~Heidrich-Meisner},
  \bibinfo{title}{{Eigenstate thermalization and quantum chaos in the Holstein
  polaron model}},
  \bibinfo{journal}{\href{http://dx.doi.org/10.1103/PhysRevB.99.155130}{Phys.
  Rev. B}} \href{http://dx.doi.org/10.1103/PhysRevB.99.155130}{{\bf
  \bibinfo{volume}{99}}, \bibinfo{pages}{155130}}
  (\href{http://dx.doi.org/10.1103/PhysRevB.99.155130}{\bibinfo{year}{2019}}).

\bibitem{iyer_oganesyan_13}
\bibinfo{author}{S.~Iyer}, \bibinfo{author}{V.~Oganesyan},
  \bibinfo{author}{G.~Refael}, and \bibinfo{author}{D.~A. Huse},
  \bibinfo{title}{Many-body localization in a quasiperiodic system},
  \bibinfo{journal}{\href{http://dx.doi.org/10.1103/PhysRevB.87.134202}{Phys.
  Rev. B}} \href{http://dx.doi.org/10.1103/PhysRevB.87.134202}{{\bf
  \bibinfo{volume}{87}}, \bibinfo{pages}{134202}}
  (\href{http://dx.doi.org/10.1103/PhysRevB.87.134202}{\bibinfo{year}{2013}}).

\bibitem{schulz_hooley_19}
\bibinfo{author}{M.~Schulz}, \bibinfo{author}{C.~A. Hooley},
  \bibinfo{author}{R.~Moessner}, and \bibinfo{author}{F.~Pollmann},
  \bibinfo{title}{Stark many-body localization},
  \bibinfo{journal}{\href{http://dx.doi.org/10.1103/PhysRevLett.122.040606}{Phys.
  Rev. Lett.}} \href{http://dx.doi.org/10.1103/PhysRevLett.122.040606}{{\bf
  \bibinfo{volume}{122}}, \bibinfo{pages}{040606}}
  (\href{http://dx.doi.org/10.1103/PhysRevLett.122.040606}{\bibinfo{year}{2019}}).

\bibitem{nieuwenburg_baum_19}
\bibinfo{author}{E.~van Nieuwenburg}, \bibinfo{author}{Y.~Baum}, and
  \bibinfo{author}{G.~Refael}, \bibinfo{title}{From bloch oscillations to
  many-body localization in clean interacting systems},
  \bibinfo{journal}{\href{http://dx.doi.org/10.1073/pnas.1819316116}{Proc.
  Natl. Acad. Sci.}} \href{http://dx.doi.org/10.1073/pnas.1819316116}{{\bf
  \bibinfo{volume}{116}}, \bibinfo{pages}{9269}}
  (\href{http://dx.doi.org/10.1073/pnas.1819316116}{\bibinfo{year}{2019}}).

\end{thebibliography}

%================================================================================
\newpage
\phantom{a}
\newpage
%%%%%%%%%%%%%%%%%%%%%%%%%%%%%%%%%%%%%%%%
\setcounter{figure}{0}
\setcounter{equation}{0}

\renewcommand{\thetable}{S\arabic{table}}
\renewcommand{\thefigure}{S\arabic{figure}}
\renewcommand{\theequation}{S\arabic{equation}}
\renewcommand{\thepage}{S\arabic{page}}

\renewcommand{\thesection}{S\arabic{section}}

\onecolumngrid

\begin{center}

{\large \bf Supplemental Material:\\
Restoring ergodicity in a strongly disordered interacting chain}\\

\vspace{0.3cm}

\setcounter{page}{1}

\  B. Krajewski$^{1}$, L. Vidmar$^{2,3}$, J. Bon\v ca$^{3,2}$, and M. Mierzejewski$^{1}$ \\
\ \\
$^1${\it Department of Theoretical Physics, Faculty of Fundamental Problems of Technology, \\ Wroc\l aw University of Science and Technology, 50-370 Wroc\l aw, Poland}\\
$^2${\it Department of Theoretical Physics, J. Stefan Institute, SI-1000 Ljubljana, Slovenia} \\
$^3${\it Department of Physics, Faculty of Mathematics and Physics, University of Ljubljana, SI-1000 Ljubljana, Slovenia} \\

\end{center}

\vspace{0.6cm}
In the Supplemental Material we provide technical details about calculations in the subspace with a fixed number of fermions,  absence of projection of $H_\Delta$ on LIOMs, the derivation of Eq.~(\ref{def_Hperpnorm}) and the lower bound on the norm of the true perturbation, the details about the nearest level spacing analysis and the fluctuations of the diagonal matrix elements. \\
\vspace{0.3cm}

\twocolumngrid

\label{pagesupp}

\section{LIOMs in the subspace with fixed number of fermions } \label{app1}

The derivation of the true perturbation $H^{\perp}_{\Delta}$, cf.~Eqs.~(\ref{hsplit})-(\ref{hpp}) in the main text, has been carried out in the Fock space of $2^L$ many-body configurations, i.e., with a variable number of fermions.
In this case, the  occupations of LIOMs $Q_\alpha$ from Eq.~(\ref{h0diag}) in the main text are independent and their products $Q^{(2)}_{\alpha,d}$ are mutually orthogonal and normalized, i.e.,
\begin{equation}
\langle Q^{(2)}_{\alpha,d} Q^{(2)}_{\alpha',d'} \rangle=\delta_{\alpha,\alpha'} \delta_{d,d'}\;.
\end{equation}
However, the actual numerical calculations presented in the main text have been carried out in a subspace with a fixed number of fermions, $N=L/2$. Then, the LIOMs are not independent because
$\sum_{\alpha=1}^L Q_{\alpha}=0$. 
This implies that the exact LIOMs in this subspace are linear combinations of $Q_\alpha$.
Nevertheless, since the interaction $H_\Delta$ has no projection on LIOMs, see Sec.~\ref{app1b}, we focus below on two-body LIOMs $Q^{(2)}_{\alpha,d}$, which are the central object in introducing the true perturbation in Eqs.~(\ref{proj}) and~(\ref{hpp}).
In the subspace of a fixed number of fermions, the two-body LIOMs are neither traceless nor  orthonormal.  In order to apply the orthogonal projections, see Eq.~(\ref{proj}) in the main text,  one first needs to construct traceless products, $Q^{(2)}_{\alpha,d}=Q_{\alpha}Q_{\alpha+d} -{\rm const}$, where the constant is set by the condition $\langle Q^{(2)}_{\alpha,d} \rangle=0$.  Then one needs to reorthogonalize the  
set $\{Q^{(2)}_{\alpha,d}\}$.  To this end we solve the eigenproblem 
\begin{equation}
\sum_{\alpha'=1}^{L} \sum_{d'=1}^{d_{max}} \langle Q^{(2)}_{\alpha,d}  Q^{(2)}_{\alpha',d'} \rangle V_{(\alpha',d'),\gamma} = \lambda_{\gamma} V_{(\alpha,d),\gamma}
\end{equation}
for the real symmetric matrix built out of all scalar products of  $Q^{(2)}_{\alpha,d}$, i.e., we solve the eigenproblem for $\langle Q^{(2)}_{\alpha,d}  Q^{(2)}_{\alpha',d'} \rangle$. Here, $V_{(\alpha,d),\gamma}$ is an orthogonal matrix and the eigenvalues are positive, $\lambda_{\gamma} > 0$, for $d_{\max} <  L/2-1$. We introduce a new set of two-body LIOMs
\begin{equation}
q_{\gamma}^{(2)}=\sum_{\alpha=1}^{L} \sum_{d=1}^{d_{max}} \frac{1}{\sqrt{ \lambda_{\gamma}}}  V_{(\alpha,d),\gamma} Q^{(2)}_{\alpha,d} \;, \label{lintra}
\end{equation}
which are normalized and mutually orthogonal
\begin{eqnarray}
\langle q_{\gamma}^{(2)} q_{\gamma'}^{(2)} \rangle  &=& \sum_{\alpha,\alpha'=1}^{L} \sum_{d,d'=1}^{d_{max}} \frac{  V_{(\alpha,d),\gamma} \langle Q^{(2)}_{\alpha,d}  Q^{(2)}_{\alpha',d'} \rangle V_{(\alpha',d'),\gamma'}} 
{\sqrt{ \lambda_{\gamma}} \sqrt{ \lambda_{\gamma'}}} \nonumber \\
&=&  \delta_{\gamma,\gamma'} \frac{ \lambda_{\gamma'}}{\sqrt{ \lambda_{\gamma}} \sqrt{ \lambda_{\gamma'}}} = \delta_{\gamma,\gamma'}.
\end{eqnarray}
The new set of orthonormal two-body LIOMs, $\{ q_{\gamma}^{(2)} \} $,  should be used instead of $\{  Q^{(2)}_{\alpha,d} \}$ whenever the Fock space is reduced to a subspace with a fixed particle number.
We note that that the linear transformation in Eq.~(\ref{lintra}) of local  $Q^{(2)}_{\alpha,d}$ leads to local $q_{\gamma}^{(2)}$. Therefore the reorthogonalization does not spoil locality of two-body LIOMs.
 
\section{Absence of projection of $H_\Delta$ on LIOMs} \label{app1b} 

We here show that $H_{\Delta} = \Delta \sum_i N_i$ from Eq.~(\ref{hd}) in the main text has no projection on the LIOMs $Q_{\alpha}$, i.e.,
\begin{equation} \label{def_noproj}
    \langle N_i Q_\alpha \rangle = 0 \;\;\; \longrightarrow \;\;\; \langle H_\Delta Q_\alpha \rangle = 0\;.
\end{equation}
This statement is valid 
irrespectively of whether the Hilbert-Schmidt inner product $\langle ... \rangle$ in  Eq.~(\ref{def_noproj}) is calculated in the Fock space (FS) of $2^L$ basis states, or within a subspace (S) with a fixed particle number $N=L/2$ and dimension $\binom{L}{N}$. To show that we express all operators in the real-space basis and calculate the projection
\begin{equation}
\langle N_i Q_{\alpha} \rangle =\frac{1}{4} \sum_{j,l} u^{*}_{l \alpha} u_{j \alpha} \langle (2n_i-1)(2n_{i+1}-1)  (2a^{\dagger}_j a_l -\delta_{jl}) \rangle, \label{projs}
\end{equation}
where $ u_{j \alpha} = \langle j|\alpha\rangle$ are components of the single-particle wavefunction of the Anderson state $| \alpha \rangle$. The only nonzero contributions to Eq.~(\ref{projs}) come from terms corresponding to $l=j$. We note also that $(2n_i-1)^2= 1$ and that the operator  $2n_i -1$ is traceless both in FS and S, i.e., 
$\langle 2n_i-1 \rangle=0$.
Therefore, the contributions corresponding to $j=i$ or $j=i+1$ vanish. In particular, for $j=i+1$ one obtains
\begin{equation}
\langle (2n_i-1)(2n_{i+1}-1)^2 \rangle=\langle (2n_i-1)\rangle=0. \label{strace1}
\end{equation}
The remaining contributions  should be considered separately for calculations in FS and S. From now on we assume that  $j \ne i$ and $j \ne i+1$ so that all lattice indexes in Eq. (\ref{projs}) correspond to different lattice sites. The vanishing of  
$\langle N_i Q_{\alpha} \rangle$ in FS follows from the independence of occupations of different sites, 
\begin{eqnarray}
 && \langle (2n_i-1) (2n_{i+1}-1) (2n_{j}-1)   \rangle    \nonumber \\ && = \langle (2n_i-1) \rangle \langle (2n_{i+1}-1)  \rangle  \langle (2n_{j}-1)  \rangle =0.
 \end{eqnarray} 
In the subspace S, we note that $\frac{1}{L} \sum_i 2n_i$ is the identity operator, hence we use the identity 
\begin{eqnarray}
0&=& \sum_i (2n_i-1)  \sum_j (2n_j-1) \sum_l (2n_l-1) \nonumber \\
&=& \sum_{i=j=l} (2n_i-1)^2 (2n_i-1) \label{con1} \\
&&+ 3 \sum_{i=l \ne j} (2n_i-1)^2 (2n_j-1)  \label{con2}  \\
&& + \sum_{i\ne j \ne l, i \ne l} (2n_i-1)(2n_j-1)(2n_l-1).  \label{con3}
\end{eqnarray}
The contributions in Eqs.~(\ref{con1}) and~(\ref{con2}) represent traceless operators, as it follows from Eq.~(\ref{strace1}). Since $\langle (2n_i-1)(2n_j-1)(2n_l-1) \rangle$ does not depend on the lattice indexes (provided that $i,j,l$ are different) one finds that
\mbox{$\langle (2n_i-1)(2n_j-1)(2n_l-1) \rangle=0$}. Therefore, all contributions to the projection $\langle N_i Q_{\alpha} \rangle $ vanish. It holds true also within the subspace S despite $\langle N_i \rangle$ is not traceless but rather $\langle N_i \rangle = O(1/L)$.

\section{Derivation of Eq.~(\ref{def_Hperpnorm})}  \label{app1c}

The r.h.s.~of Eq.~(\ref{def_Hperpnorm}) in the main text corresponds to the norm of of the difference
\begin{equation} \label{def_norm_difference}
    || H_\Delta - H_\Delta^\parallel ||^2 =
    \langle (H_\Delta - H_\Delta^\parallel) (H_\Delta - H_\Delta^\parallel) \rangle \;.
\end{equation}
Equation~(\ref{hsplit}) implies that $\langle H_\Delta H_\Delta^\parallel \rangle = \langle H_\Delta^\parallel{}^2 \rangle $, hence one can rewrite Eq.~(\ref{def_norm_difference}) as 
\begin{equation} \label{def_norm_difference_2}
    || H_\Delta - H_\Delta^\parallel ||^2 =
    \langle H_\Delta ^2 \rangle - \langle H_\Delta H_\Delta^\parallel \rangle \;.
\end{equation}
Since $H_\Delta^\parallel = \Delta \sum_i N_i^\parallel$ and using the expression in Eq.~(\ref{proj}) for $N_i^\parallel$, one obtains
\begin{align} \label{def_norm_difference_3}
    \langle H_\Delta H_\Delta^\parallel \rangle & = \Delta \sum_i \sum_{d,\alpha}
    \langle Q_{\alpha,d}^{(2)} N_i \rangle \langle H_\Delta Q_{\alpha,d}^{(2)} \rangle  = \sum_{d,\alpha} \langle H_\Delta Q_{\alpha,d}^{(2)}\rangle^2.
\end{align}
Pluging  Eq.~(\ref{def_norm_difference_3}) into Eq.~(\ref{def_norm_difference_2}), one obtains the expression on the r.h.s.~of Eq.~(\ref{def_Hperpnorm}) in the main text.

\section{Bounds for projected operators}  \label{app2}

Here, we study  in more details  the properties of the projected operators, $N^{\perp}_i$, and establish a lower bound on their norm.
To this end we use the many-body Anderson states $| \vec{\alpha} \rangle=|\alpha_1,\alpha_2,... ,\alpha_L \rangle$ and introduce an auxiliary  operator
\begin{equation}
\tilde{N}_i=N_i-\sum_{\vec{\alpha}} \langle \vec{\alpha}|N_i|  \vec{\alpha} \rangle \; |\vec{\alpha} \rangle \langle \vec{\alpha} |\;, \label{fulproj}
\end{equation}
where all diagonal matrix elements have been eliminated. We note that the projection in Eq.~(\ref{proj}) in the main text eliminates the diagonal matrix elements of $N_i^{\perp}$ only partially, hence it is intuitively clear that 
$ ||N_i^{\perp} || \ge || \tilde{N}_i||  $. Below we present a formal proof of this lower bound on $ ||N_i^{\perp} ||$.  

We rewrite Eq.~(\ref{proj}) in the main text as
\begin{equation}
N_i=N^{\perp}_i + \sum_{\beta,d}  \langle Q^{(2)}_{\beta,d} N_i \rangle   Q^{(2)}_{\beta,d} \label{ss1}
\end{equation} 
and note that $Q^{(2)}_{\beta,d}$ are diagonal in the many-body Anderson basis,  hence
\begin{equation} 
\sum_{\vec{\alpha}} \langle \vec{\alpha}| Q^{(2)}_{\beta,d}  |  \vec{\alpha} \rangle \; |\vec{\alpha} \rangle \langle \vec{\alpha}| =Q^{(2)}_{\beta,d} \;. \label{ss2}
\end{equation} 
Putting  Eqs.~(\ref{ss1}) and~(\ref{ss2})   into the right-hand side of  Eq.~(\ref{fulproj}) one obtains the identity
\begin{equation}
N^{\perp}_i  = \tilde{N}_i   + \sum_{\vec{\alpha}} \langle \vec{\alpha}|N^{\perp}_i    |  \vec{\alpha} \rangle \; |\vec{\alpha} \rangle \langle \vec{\alpha}|. \label{ss3}
\end{equation}
In the  Anderson basis, $ \tilde{N}_i$  has only off-diagonal matrix elements whereas the second term, $ \sum_{\vec{\alpha}} \langle \vec{\alpha}|N^{\perp}_i    |  \vec{\alpha} \rangle \; |\vec{\alpha} \rangle \langle \vec{\alpha}|$, is 
diagonal. Moreover, the squared Hilbert-Schmidt norm can be explicitly written as a sum of squares  of all matrix elements,  $ ||...||^2= \frac{1}{Z} \sum_{\vec{\alpha},\vec{\alpha}'} |  \langle \vec{\alpha}| ...    |  \vec{\alpha}' \rangle|^2$. The latter two properties combined with Eq.~(\ref{ss3})
imply that
\begin{equation}
||N^{\perp}_i ||^2=||  \tilde{N}_i  ||^2+   || \sum_{\vec{\alpha}} \langle \vec{\alpha}|N^{\perp}_i    |  \vec{\alpha} \rangle \; |\vec{\alpha} \rangle \langle \vec{\alpha}| \; ||^2 \ge ||  \tilde{N}_i  ||^2.
\end{equation} 
Consequently, one obtains the lower bound 
\begin{equation}
||N^{\perp}_i ||^2  \ge ||  \tilde{N}_i  ||^2, \label{bound0}
\end{equation}
 which holds true for arbitrary distance $d_{max}$ in Eq. (\ref{proj}) in the main text.

\begin{figure}[!b]
\includegraphics[width=1.0\columnwidth]{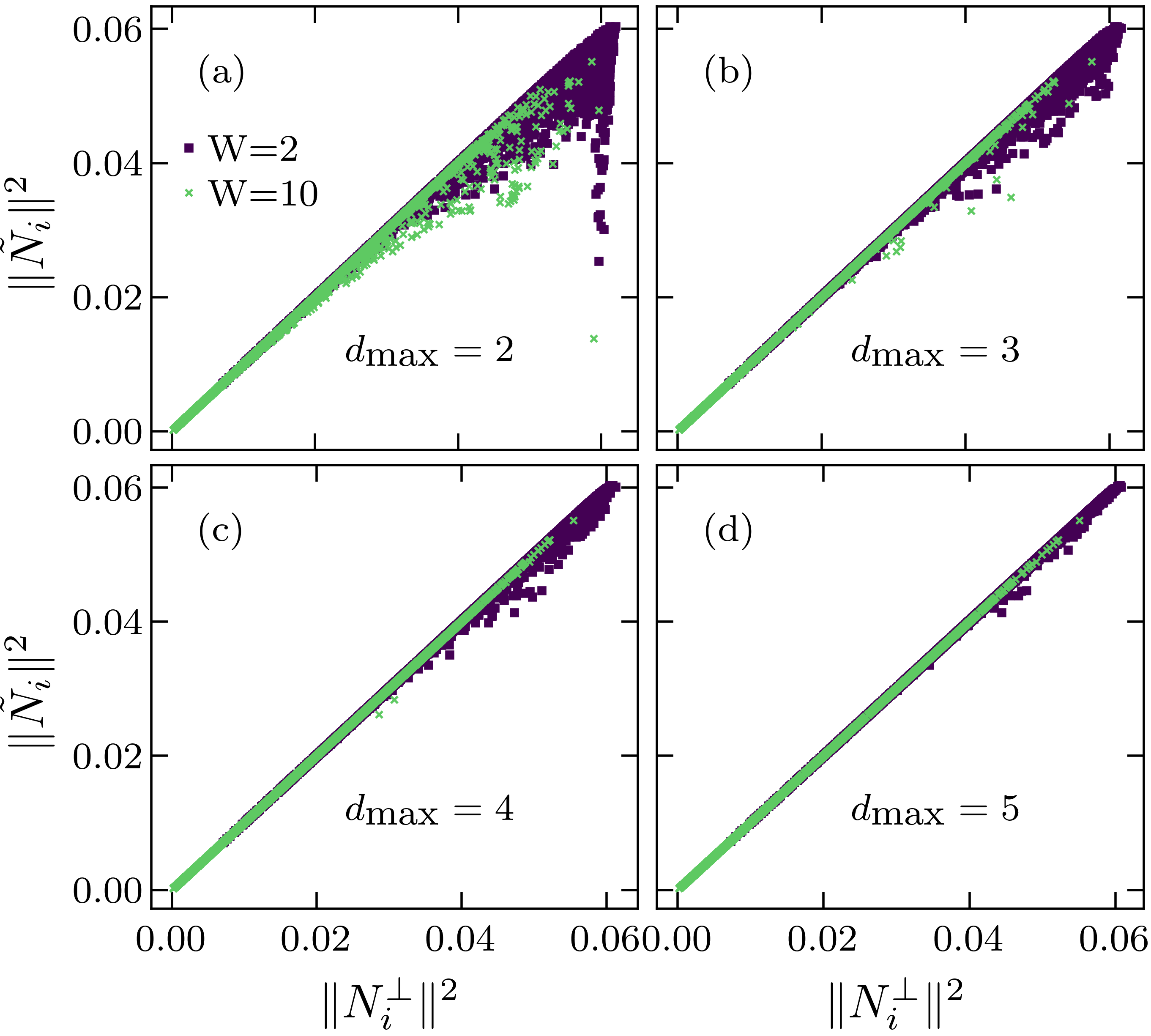}
\caption{Results for chain with $L=14$ sites and $L/2$ fermions. Points show $||N_i^{\perp} ||$ and $||\tilde{N}_i || $   for various sites $i$, disorder realizations and $W$. In (a),  (b), (c) and (d)  we use, respectively, the maximal distance
 $d_{max}=2,3,4$ and 5, see Eq. (\ref{proj}) in the main text.
}
\label{figs1}
\end{figure}

Figure~\ref{figs1} shows  correlations between norms of $N_i^{\perp}$ and  $ \tilde{N}_i$ for various $d_{max}$. Each point in this plot shows result for a single site $i$ and a single realization of  disorder. These results not only confirm the
bound  $ ||N_i^{\perp} || \ge || \tilde{N}_i||$ from Eq.~(\ref{bound0}) but also demonstrate that for strong disorder (i.e., for small $||N_i^{\perp} ||$) the projected operator can be well approximated as $N_i^{\perp}  \simeq \tilde{N}_i$. Comparing Figs.~\ref{figs1}(a)-\ref{figs1}(d) one observes  that the larger is the distance  $d_{max}$, the better is the approximation.
At strong disorder, however, the approximation $N_i^{\perp}  \simeq \tilde{N}_i$ is accurate already for  $d_{max}=2$.  It means that the conserved part of the interaction, $N_i$, is dominated by the projections of $N_i$ on two-body LIOMs $Q^{(2)}_{\alpha,d}$, i.e., by the products of LIOMs corresponding to the neighboring orbitals with $d=1$ or $d=2$.

Next, we discuss the origin of the bound  $|| N^{\perp}_i ||\ge 1/(8W)^2$ which was observed from the numerical results shown in the main text in Fig. \ref{fig1}.  To this end we show that $|| \tilde{N}_i ||\ge 1/(8W)^2$ and
then make use of the inequality  (\ref{bound0}). For simplicity we assume that the single-particle  wave functions, $u_{i \alpha}$, are real,
we express the operator $N_i$ in the Anderson basis
\begin{eqnarray}
&& N_i =  \left(a^{\dagger}_{i}  a_{i}- \frac{1}{2}\right) \left(a^{\dagger}_{i+1}  a_{i+1}-\frac{1}{2}\right) \label{nia} \\ 
&&=\sum_{\alpha,\beta,\gamma,\eta} u_{i \alpha}   u_{i \beta} (a^{\dagger}_{\alpha}  a_{\beta}-\frac{\delta_{\alpha,\beta}}{2})  u_{i+1 \gamma}   u_{i+1 \eta} (a^{\dagger}_{\gamma}  a_{\eta}-\frac{\delta_{\gamma,\eta}}{2}) \;,  \nonumber
\end{eqnarray} 
and  recall that $\tilde{N}_i$ represents the off-diagonal contribution to Eq.~(\ref{nia}).
The more localized are the Anderson wave-functions the smaller are the off-diagonal contributions to Eq.~(\ref{nia}).
Then, it is useful to study a two-site problem for the Anderson Hamiltonian [see Eq.~(\ref{h0}) in the main text]  with extreme values of the disorder potentials $\epsilon_{1,2}=\pm W$
 \begin{equation}
 \left(
 \begin{array} {cc}
 W & \frac{1}{2} \\
  \frac{1}{2} & -W \\
 \end{array} 
 \right)
 \left(
  \begin{array} {c}
u_{1 \alpha} \\
u_{2 \alpha} \\
 \end{array} 
 \right)
 =\varepsilon_{\alpha}
\left(
  \begin{array} {c}
u_{1 \alpha} \\
u_{2 \alpha} \\
 \end{array} 
 \right). 
  \label{sw2}
 \end{equation}  
Direct calculations show that $\lim_{W \to  \infty} u_{11}=\lim_{W \to  \infty} u_{22}=1$,  whereas, 
\begin{equation}
\lim_{W \to  \infty} W u_{i \alpha }= \pm \frac{1}{4}, \quad \quad i \ne \alpha. \label{s2}
\end{equation}
We assume that the lower bound on    $|| \tilde{N}_i || ^2$ denoted as $ ||N^{\rm bound}_i ||^2$,  can be obtained via introducing single-particle wave-functions from Eq.~(\ref{sw2}) into  Eq.~(\ref{nia}). 
In other words,  $N^{\rm bound}_i $ corresponds to the most localized orbitals, $u_{i \alpha}$, on two sites.
Namely, we assume that for each site $i$ there is a single Anderson state denoted as $\alpha(i)$
such that $u_{i \alpha(i)}$ is of the order $O(1)$, and one other state, $\alpha'(i) \ne \alpha(i)$, for which $u_{i \alpha'(i)}=\pm \frac{1}{4W} $,
\begin{equation}
u_{i \beta} \simeq  \delta_{\beta,\alpha(i)} \pm \frac{1}{4W}   \delta_{\beta,\alpha'(i)}. \label{sup11}
\end{equation}
Then, one  finds the leading  (with respect to $1/W$) contributions to the off-diagonal part of $N^{\rm bound}_i $,
 \begin{eqnarray}
 N^{\rm bound}_i &\simeq & \pm  \frac{1}{4W} [a^{\dagger}_{\alpha'(i)}a_{\alpha(i)} + {\rm H.c.}][n_{\alpha(i+1)}-\frac{1}{2}] \nonumber \\
 && \pm  \frac{1}{4W}[n_{\alpha(i)}-\frac{1}{2}] [a^{\dagger}_{\alpha'(i+1)}a_{\alpha(i+1)} + {\rm H.c.}]\;. \nonumber \\ \label{bound1}
 \end{eqnarray}
Using the identity $[n_{\alpha(i)}-\frac{1}{2}] | \vec{\alpha} \rangle =\pm \frac{1}{2} | \vec{\alpha} \rangle $ one may  simplify Eq.~(\ref{bound1}) to
  \begin{eqnarray}
 N^{\rm bound}_i &\simeq & \pm  \frac{1}{8W} [a^{\dagger}_{\alpha'(i)}a_{\alpha(i)} + {\rm H.c.}] \nonumber \\
 && \pm  \frac{1}{8W} [a^{\dagger}_{\alpha'(i+1)}a_{\alpha(i+1)} + {\rm H.c.}]. \nonumber \\ \label{bound2}.
 \end{eqnarray}
The resulting  $N^{\rm bound}_i$  is a sum of two hopping terms and the squared Hilbert-Schmidt norm of each term equals $\frac{1}{2} \frac{1}{(8W)^2} $. Finally, we find  the inequalities
\begin{equation}
   ||N^{\perp}_i ||^2 \ge   ||\tilde{N}_i ||^2 \ge   ||N^{\rm bound}_i ||^2=1/(8W)^2.
 \end{equation}  
Fig.~\ref{fig1} in the main text demonstrates that $1/(8W)^2$ very accurately reproduces the minimum of  $||N^{\perp}_i ||^2 $ obtained from  numerical simulations 
already for $W>3$. It follows from Eq.~(\ref{sup11}) that the latter bound is applicable only at $1/4W \ll 1$ and it must break down at $W<0.5$ since
$||N^{\perp}_i ||^2 \le ||N_i ||^2 =1/16.$

\begin{figure}[!]
\includegraphics[width=1.0\columnwidth]{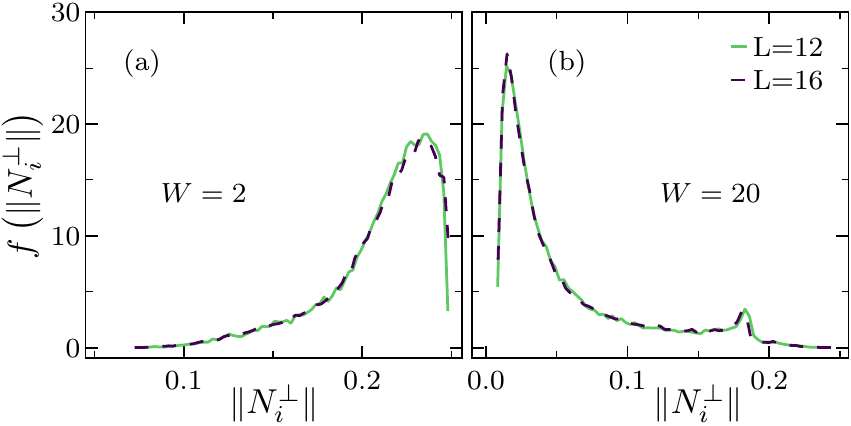}
\caption{Probability density functions $f$ of $||N^{\perp}_i ||$. Results are obtained for 4000 realizations of disorder at $L=12,16$ with (a) $W=2$ and (b) $W=20$.}
\label{figs2}
\end{figure}

Finally we demonstrate that the norms $|| N^{\perp}_i||$, shown in Fig.~\ref{fig1} in the main text, weakly depend on the size of the studied system. To this end,
we have determined the distributions of the latter quantities for lattices with $L=12$ and $L=16$ sites.  
Figures~\ref{figs2}(a) and \ref{figs2}(b) show the probability density functions of $||N^{\perp}_i ||$ at $W=2$ and $W=20$, respectively. The distributions have been obtained via collecting results for 4000 realizations of disorder and for all sites, $i=1,...,L$. We do not observe any significant $L$-dependence of these distributions for either weak  (left panel) or strong (right panel) disorder.

\section{Nearest level spacings}  \label{app3}

In Fig.~\ref{fig2} in the main text we studied the statistical properties of the ratio of nearest level spacings, shortly the gap ratio~\cite{oganesyan07}.
For a target many-body eigenstate $|E_m\rangle$, the gap ratio is defined as
\begin{equation} \label{def_rm}
    r_m = \frac{{\rm min}\{\delta E_m, \delta E_{m-1}\}}{{\rm max}\{ \delta E_m, \delta E_{m-1} \}} \;,
\end{equation}
where $\delta E_m = E_{m+1} - E_m$ is the level spacing.
We then averaged $r_m$ over the middle third of the energy spectrum as well as over 4000 various realizations of disordered. The average gap ratio $\langle r \rangle$ is plotted in the main panels of Figs.~\ref{fig2}(a) and~\ref{fig2}(b).

In the inset of Fig.~\ref{fig2}(b) we plotted a probability density function $P(r)$ that includes results for $r_m$ from Eq.~(\ref{def_rm}) obtained at different disorder realizations, as well as different target eigenstates $|E_m\rangle$ at a fixed disorder realization.
The latter are again obtained from the middle third of the energy spectra.
Results are compared to the analytical predictions for the Poisson distribution~\cite{oganesyan07}, 
\begin{eqnarray}
    P(r) = \frac{2}{(1+r)^2} \;,
\end{eqnarray}
see the dash-dotted line the inset of Fig.~\ref{fig2}(b), and for the GOE~\cite{atas2013},
\begin{equation}
    P(r) = \frac{27}{4} \frac{r + r^2}{(1+r+r^2)^{5/2}} \;,
\end{equation}
see the dashed line in the inset of Fig.~\ref{fig2}(b).

\begin{figure}[!]
\includegraphics[width=1.0\columnwidth]{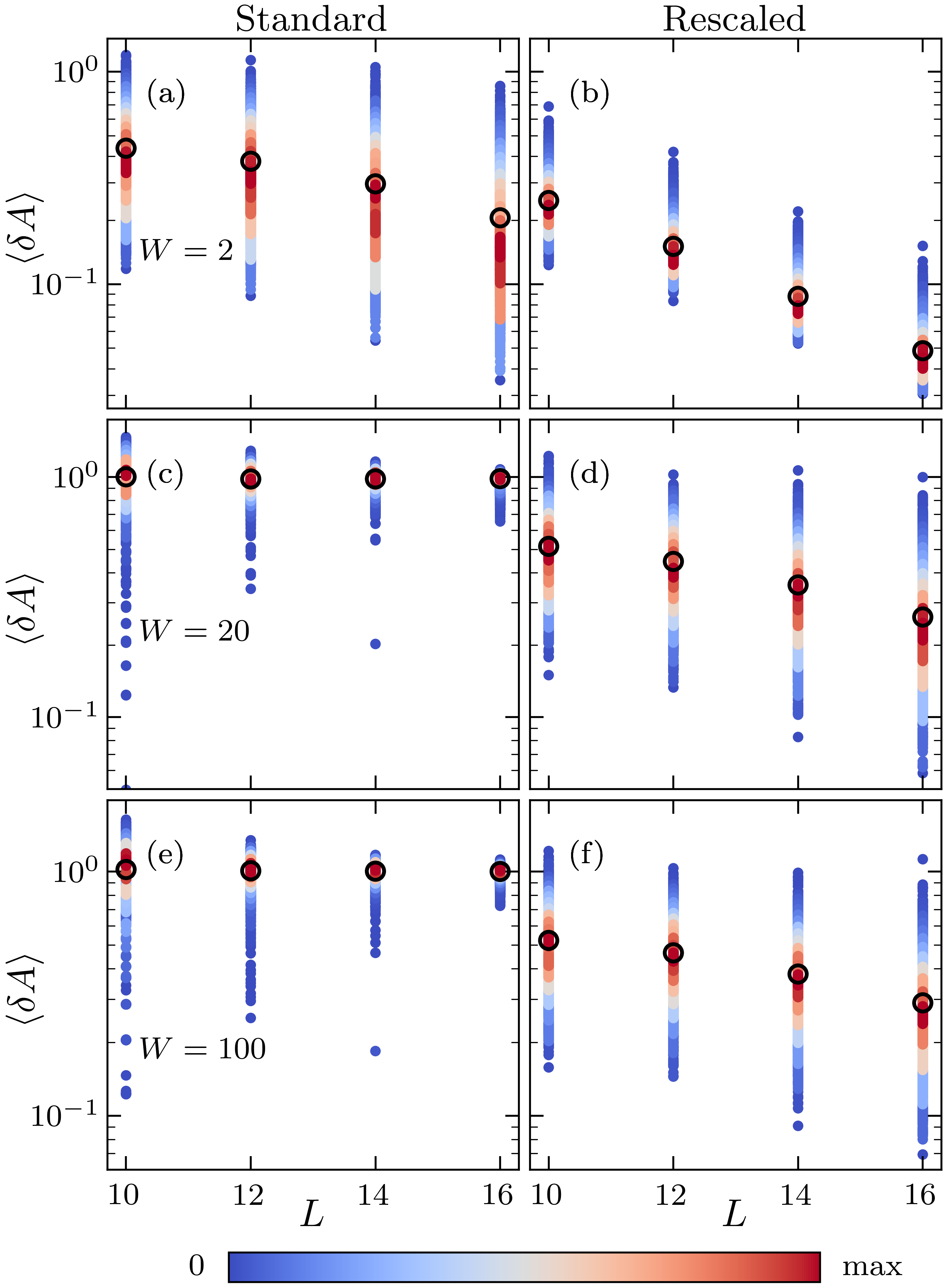}
\caption{Eigenstate-to-eigenstate fluctuations $\langle \delta A \rangle$ from Eq.~(\ref{eefluct}) in the main text.
We first calculate $\langle \delta A \rangle$ for a single lattice occupation, $A=2n_i-1$, and different realizations of disorder. Then present a density plot of these values. Open circles show the medians.
Panels (a,c,e) show results for the standard model, $H$, and (b,d,f) for the rescaled model $\tilde{H}$.}
\label{figs3}
\end{figure}

\section{Matrix elements of observables}  \label{app4}

In the main text we studied  eigenstate-to-eigenstate fluctuations of the diagonal matrix elements, $\langle \delta A \rangle$, for a  single site occupation, $A=2n_i-1$. In particular, we have determined
the probability density functions, $f(\langle \delta A \rangle)$ shown in Fig.~\ref{fig4} in the main text. In order to better visualize the $L$-dependence of the eigenstate-to-eigenstate fluctuations,
we show in Figure~\ref{figs3}  the density plot of $\langle \delta A \rangle$ as a function of the system size. 
Each point shows result for a single site, $i$, and a single realization of disorder whereas color marks the density of such points.
In the standard model at strong disorder, see Figs.~\ref{figs3}(c) and~\ref{figs3}(e), one observes strong fluctuations and $\langle \delta A \rangle) \sim 1$ for all accessible system sizes.
In the rescaled model, see Figs.~\ref{figs3}(d) and~\ref{figs3}(f), the fluctuations visibly decrease with $L$, however the probability density function, $f(\langle \delta A \rangle)$,
remains broad. Therefore, the $L$-dependence of the fluctuations can be followed via inspecting the medians (circles) and maxima (red color) of $f(\langle \delta A \rangle)$. Both quantities suggest that fluctuations decay exponentially with $L$ as it is expected for systems which obey the eigenstate thermalization
hypothesis.

%===================================
%=============================================
\end{document}